\documentclass[a4paper,11pt]{article}

\usepackage{jheppub}

\usepackage{placeins}
\newcounter{jheppanel}[figure]
\renewcommand{\thejheppanel}{\alph{jheppanel}}

\title{Improved method for extracting low-energy expansion from OPE in mass-gapped theories}
\author[a]{Hajime Furuta}
\author[a]{and Yukinari Sumino}

\affiliation[a]{Department of Physics, Tohoku University, \\
Sendai, 980-8578 Japan}
\emailAdd{furuta.hajime.r1@dc.tohoku.ac.jp}
\emailAdd{yukinari.sumino.a4@tohoku.ac.jp}

\abstract{Recently, Takaura proposed a new method for extracting the first nonzero coefficient of the low-energy expansion of an observable from its operator product expansion (OPE) in asymptotically free theories with a mass gap and a single threshold. The method uses a weight function to suppress the contribution of the ``duality-violation'' (DV) term that arises when relating the OPE to the low-energy expansion. We extend this method to low-energy expansion coefficients at arbitrary orders. We present two approaches: Method I uses weight functions that isolate the coefficients of low-energy expansion from the OPE. Method II subtracts the DV contribution from the OPE before the extraction. Applying these methods to the two-dimensional large-$N$ $O(N)$ nonlinear sigma model, we find that both methods successfully reconstruct higher-order low-energy expansion coefficients.}

\preprint{TU-1332}
\begin{document}
\maketitle
\flushbottom

\section{Introduction}

In asymptotically free theories such as QCD, the running coupling constant becomes small at high energies and large at low energies. As a consequence, the perturbative expansion in the coupling constant, which is a standard tool in quantum field theory (QFT), is not applicable at low energies. At high energies, on the other hand, perturbation theory is applicable. Furthermore, the operator product expansion (OPE), which is a double expansion in inverse powers of the energy scale and in the coupling constant, can describe physical observables including non-perturbative effects. 

Unlike at high energies, no general systematic framework analogous to the OPE is available for generic observables in the low-energy region. Takaura~\cite{takaura2024} proposed a method to obtain the low-energy limit from the OPE of an observable with a mass gap by exploiting its analytic properties. The method considers one-scale observables with a single threshold and starts from the OPE, including the perturbatively calculated Wilson coefficients and non-perturbative matrix elements. Its aim is to extract coefficients of the low-energy expansion from this high-energy information.

If an observable $S$ has a mass gap, it is analytic in the region $|p^2|<M_{\mathrm{gap}}^2$, where $p$ is the external momentum. Therefore, in principle, it admits a Taylor expansion around $p^2 = 0$, namely, a low-energy expansion. Takaura showed that at least its first nonzero coefficient can be extracted from the OPE through the following procedure:

\begin{enumerate}
  \item Consider a ``Borel transform'' of the low-energy expansion: $S_{\mathrm{low}}(p^2) \rightarrow \tilde{S}_{\mathrm{low}}(z)$.
  \item Apply the same transform to the OPE:\footnotemark{} $S_{\mathrm{OPE}}(p^2) \rightarrow \tilde{S}_{\mathrm{OPE}}(z)$.
  \item Extract the low-energy expansion coefficient using the Cauchy integral formula,
  while suppressing the effect of the duality violation (DV) term generated by the transform.
\end{enumerate}

\footnotetext{This transform was introduced in ref.~\cite{Hayashi2023} and is called the ``dual transform''. However, the transform was originally intended to suppress IR renormalons, which is different from the purpose considered here.}

Here, when referring to a ``Borel transform,'' it is important to specify which series the transform is applied to. Unless otherwise stated, the Borel transform considered in ref.~\cite{takaura2024} and mainly used in this paper refers to the transform of the low-energy expansion in positive powers of the momentum, which divides its coefficients by the factorial of the order: $(p^2)^m \to (p^2)^m /m!$. The Borel transform often used in QCD sum rules~\cite{SHIFMAN1979385} divides the coefficients of the OPE in inverse powers of the momentum by the factorial of the order: $(1/p^2)^n \to (1/p^2)^n /n!$.\footnotemark{} The Borel transform used to analyze divergences associated with renormalons and instantons in perturbation theory~\cite{tHooft1977} divides the coefficients of the expansion in powers of the coupling constant by the factorial of the order: $g^k \to g^k/k!$. 

\footnotetext{QCD sum rules and the method of ref.~\cite{takaura2024}, extended here, differ not only in their Borel transforms but also in their aims. The former relate the OPE to (experimental or modeled) spectral integrals primarily to determine hadronic and QCD parameters, whereas the latter extracts low-energy expansion coefficients from the OPE, including non-perturbative matrix elements.}

The Borel transform of the low-energy expansion extends the finite radius of convergence of the low-energy expansion determined by the mass gap. Then $\tilde{S}_{\mathrm{low}}(z)$ converges for arbitrarily large $z$, and the contour of the Cauchy integral for extracting the low-energy expansion coefficients can be enlarged accordingly.

On the other hand, the original OPE is valid in the high-energy region. One might therefore naively expect that its Borel transform $\tilde{S}_{\mathrm{OPE}}(z)$ also provides a good approximation to $\tilde{S}_{\mathrm{low}}(z)$ along a Cauchy integration contour of sufficiently large radius in the $z$ plane.

However, even at large $|z|$, there are regions where $\tilde{S}_{\mathrm{OPE}}(z)$ differs significantly from $\tilde{S}_{\mathrm{low}}(z)$. Takaura refers to this mismatch as a ``duality-violation'' (DV) effect.\footnotemark{} 
\footnotetext{The term ``DV'' is also commonly used for quark-hadron duality violations~\cite{SHIFMAN_2001}, where small contributions not captured by the Euclidean OPE can become relevant after analytic continuation to the Minkowski region. The DV considered here also becomes prominent in the Minkowski region, but has a different origin: it arises because the transformation of the OPE picks up contributions from the low-energy region where the OPE is not valid.}
Following Takaura, we express this mismatch as $\tilde{S}_{\mathrm{DV}}(z)=\tilde{S}_{\mathrm{OPE}}(z)-\tilde{S}_{\mathrm{low}}(z)$.\footnotemark{}
\footnotetext{The sign convention for $\tilde{S}_{\mathrm{DV}}$ originally in ref.~\cite{takaura2024} is opposite to ours.}

Therefore, the DV contribution must be removed or suppressed to extract the low-energy expansion coefficients. Takaura proposed to use a weight function that suppresses the DV contribution from the region where the DV is large. However, an explicit construction was given only for the first nonzero coefficient of the low-energy expansion.

In this study, we develop two methods for extracting low-energy expansion coefficients at arbitrary orders from the OPE, as an extension of the previous work. We consider the following two approaches, referred to as Method I and Method II\@. In Method I, we construct a weight function with two properties: when applied to $\tilde{S}_{\mathrm{low}}$, the weighted contour integral extracts the target coefficient $c_m$ without mixing other low-energy coefficients; when applied to the asymptotic expansion of $\tilde{S}_{\mathrm{DV}}$, it makes the contribution from each term vanish. In Method II, we subtract the DV contribution when performing the transform in step 2.

The basic idea of Method II was already pointed out in ref.~\cite{takaura2024}, but here we construct the procedure explicitly. We also compare how the results of the previous method and Methods I and II depend on the OPE truncation order.

The fact that the low-energy expansion can be obtained at arbitrary orders further supports the conjecture in ref.~\cite{takaura2024} that the OPE, although formulated as a high-energy expansion, contains information about the low-energy region. 

In section 2, we review the relation among the low-energy expansion, the OPE and the DV term established in ref.~\cite{takaura2024}, as well as the method for extracting the low-energy limit. In section 3, we propose two formulas for extracting low-energy expansion coefficients at arbitrary orders as extensions of the previous method. In section 4, we apply these methods to the two-dimensional large-$N$ $O(N)$ nonlinear sigma model and test their validity by comparing the extracted low-energy expansion with the exact result. This model is exactly solvable and is an asymptotically free theory with a mass gap, and has therefore often been studied as a toy model of QCD~\cite{Beneke1998,Beneke1999,NOVIKOV1984103,DAVID1982433}. In appendix A, we derive explicit expressions for the transformed OPE and DV contributions and their difference, and discuss the effects of anomalous dimensions. In appendix B, we analytically evaluate the additional threshold contribution arising from the $\sigma$-field self-energy correction and discuss its treatment in the extraction of low-energy expansion coefficients.

\section{Brief review of previous work}\label{sec:review}

In this section, we review the parts of the method proposed in ref.~\cite{takaura2024} that are relevant to the present study. We first summarize the properties of the quantities considered in this work. 

In an asymptotically free theory with a mass gap, we consider an observable $S(p^2)$ satisfying the following properties:
\begin{itemize}
  \item[a.] Renormalization-group invariant and depends on a single external momentum scale $p^2$, where $p$ has mass dimension 1.
  \item[b.] The high-energy expansion in terms of the OPE is available.
  \item[c.] Only one threshold exists, located at $p^2 = -M_{\mathrm{gap}}^2$.
\end{itemize}

Here and throughout this paper, we use the Euclidean metric. The value of $M_{\mathrm{gap}}$ does not necessarily have to be given as an input; it can be estimated from the OPE in the course of the calculation (see section~\ref{sec:StildeASYM}).

The OPE is a double expansion in the coupling constant $g(\mu^2)$ and the inverse squared external momentum $1/p^2$ of the following form:
\begin{align}
  S_{\mathrm{OPE}}(p^2) &= (p^2)^j \left( C_{0,\mathrm{PT}}(p^2, \mu^2) \langle \mathcal{O}_0 \rangle + C_{1,\mathrm{PT}}(p^2, \mu^2) \frac{\langle \mathcal{O}_1 \rangle_{\mu^2}}{(p^2)^{\frac{d_1}{2}}} + \dots \right) \label{eq:OPE}\\
  &= (p^2)^j \sum_{n=0}^{\infty} C_{n,\mathrm{PT}}(p^2, \mu^2) \frac{\langle \mathcal{O}_n \rangle_{\mu^2}}{(p^2)^{\frac{d_n}{2}}}, \label{SOPE}
\end{align}
\begin{align}
  C_{n,\mathrm{PT}}(p^2, \mu^2) = \sum_{k=0}^{\infty} d_{n,k}\left( \log \frac{\mu^2}{p^2} \right) g(\mu^2)^{k+1}. \label{eq:WilsonPT}
\end{align}
Here, $C_{n,\mathrm{PT}}$ is the Wilson coefficient. $\langle \mathcal{O}_n \rangle_{\mu^2}$ is a non-perturbative matrix element of mass dimension $d_n$, which can be determined from experiments or lattice calculations. $\mu$ is an arbitrary renormalization scale, and $d_{n,k}$ is a polynomial in $\log(\frac{\mu^2}{p^2})$. The coupling constant behaves as $g(\mu^2)\sim 1/\log(\mu^2/M^2)$,\footnotemark{} where $M$ is a characteristic mass scale of the theory. The dimension of the observable $S$ is the same as $p^{2j}$, where $j$ is an integer. In the following, we assume $d_n/2=n$ for simplicity. We also assume that $\langle \mathcal{O}_n \rangle_{\mu^2}$ is independent of the renormalization scale, although the discussion can be extended to the case with anomalous dimensions (see appendix~\ref{sec:anomalous}).

\footnotetext{In the two-dimensional $O(N)$ nonlinear sigma model considered in the latter part of this paper, the coupling constant is exactly given by $g(\mu^2)=1/\log(\mu^2/M^2)$. This simplifies the numerical calculations, but the conclusions of this paper remain valid even when the coupling constant does not take this form.}

The OPE is valid for $|p^2| > M_{\mathrm{gap}}^2$ and can in principle be calculated order by order starting from the lowest order. However, the perturbative expansion of each Wilson coefficient $C_{n,\mathrm{PT}}$ is generally asymptotic and diverges as the truncation order $k$ tends to infinity~\cite{tHooft1977}. When the divergence of $C_{n,\mathrm{PT}}$ originates from renormalons, it is believed that a finite result for the physical quantity can be obtained by absorbing the ambiguity associated with the divergence into the non-perturbative matrix elements~\cite{MUELLER1985327}. On the other hand, apart from the divergence of the expansion of the Wilson coefficients in the coupling constant, the expansion of $S_{\mathrm{OPE}}(p^2)$ in $1/p^2$ with respect to $n$ is expected to converge for $|p^2| > M_{\mathrm{gap}}^2$~\cite{Beneke1998}.\footnotemark{} In practical calculations, the expansions are truncated at finite orders $k=k_{\mathrm{max}}$ and $n=n_{\mathrm{max}}$.

\footnotetext{This is because the singularity nearest to $p^2=\infty$ for the expansion in $1/p^2$, i.e., the expansion around $p^2=\infty$, is the threshold at $p^2=-M_{\mathrm{gap}}^2$. This argument applies only when the number of particle thresholds is finite. In theories with infinitely many particle thresholds extending to $p^2=\infty$, as in theories at finite $N$, the radius of convergence is expected to be zero.}

On the other hand, we express the low-energy expansion of $S(p^2)$ as
\begin{align}
  S_{\mathrm{low}}(p^2) = M^{2j}\sum_{m=0}^{\infty}c_m \left(\frac{p^2}{M^2}\right)^m. \label{eq:LEE}
\end{align}
The analyticity around the origin in the $p^2$ plane implies that this expansion converges to the exact value of $S(p^2)$ for $|p^2| < M_{\mathrm{gap}}^2$. However, calculating $c_m$ directly is generally difficult.

\subsection{Relation between OPE and low-energy expansion}\label{sec:relation}

\begin{figure}
\centering
\begin{minipage}[t]{0.47\linewidth}
  \centering
  \includegraphics[width=\linewidth, trim=10mm 100mm 10mm 40mm,clip]{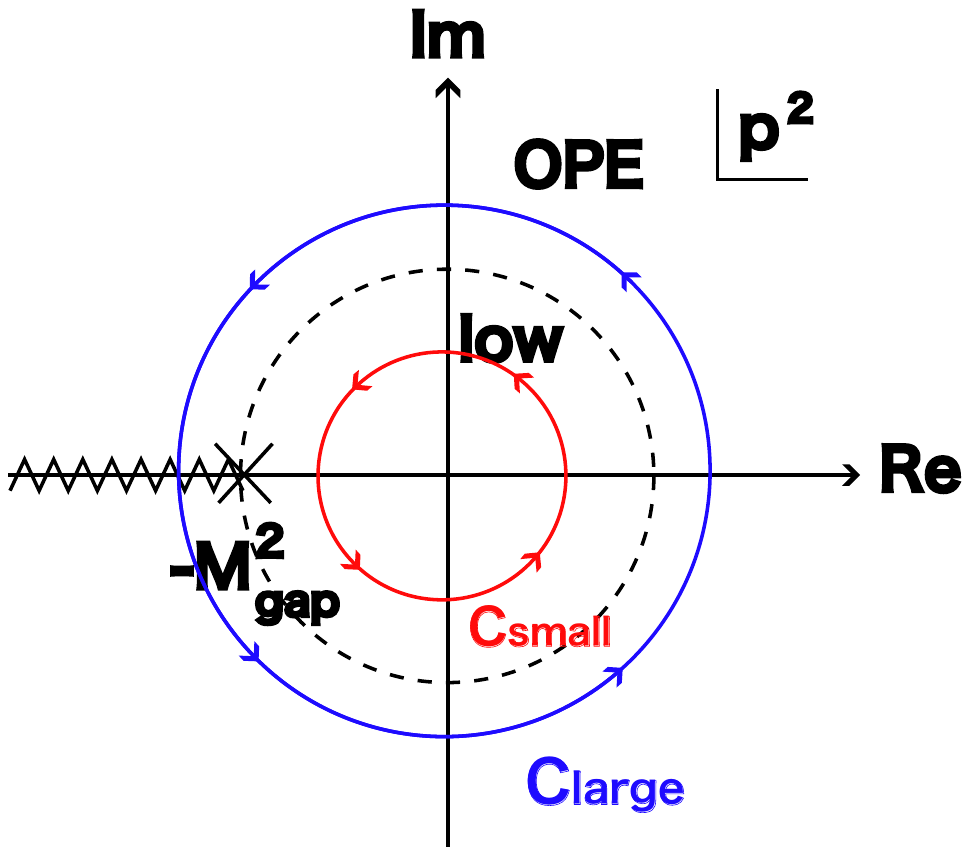}
  \par\smallskip
  \refstepcounter{jheppanel}\label{beforeborel}(\thejheppanel)
\end{minipage}
\hfill
\begin{minipage}[t]{0.47\linewidth}
  \centering
  \includegraphics[width=\linewidth]{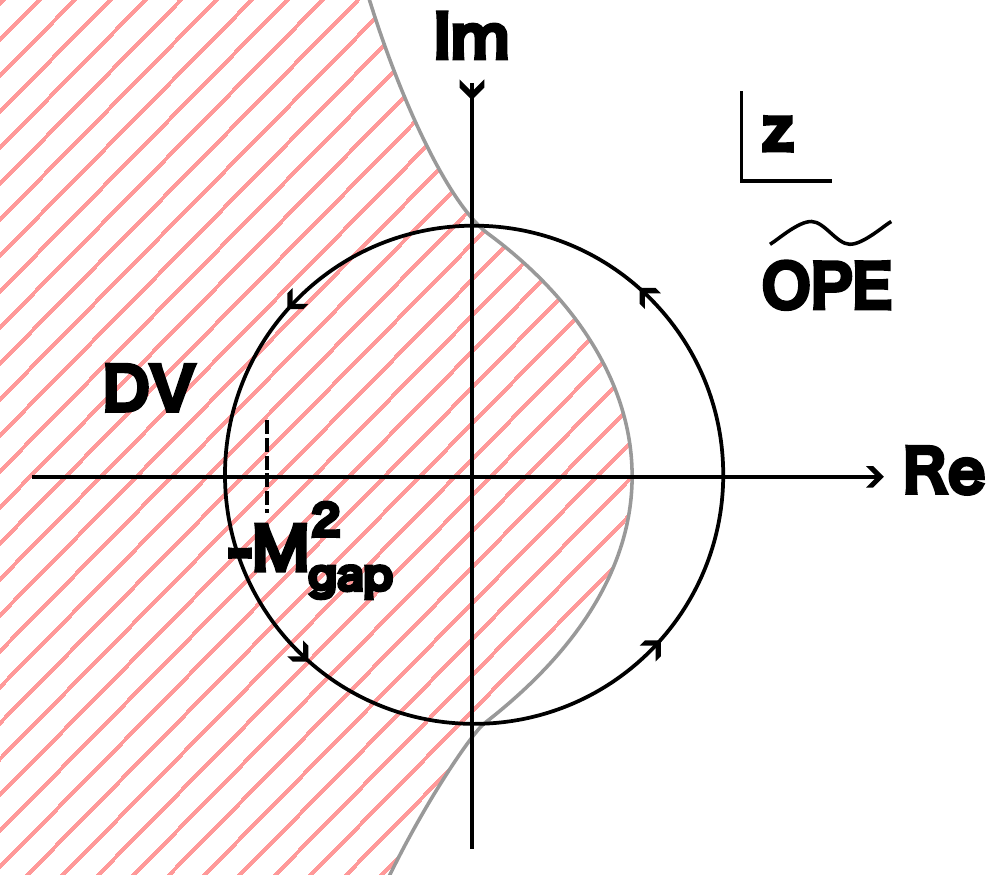}
  \par\smallskip
  \refstepcounter{jheppanel}\label{afterborel}(\thejheppanel)
\end{minipage}
\caption{Analytic structures before and after the Borel transform.
(a) Analytic structure of $S_{\mathrm{OPE}}(p^2)$ and $S_{\mathrm{low}}(p^2)$ before the Borel transform. $S_{\mathrm{low}}$ converges only for $|p^2| < M_{\mathrm{gap}}^2$, hence the integration contour $C_{\mathrm{small}}$ cannot be deformed into $C_{\mathrm{large}}$, on which $S_{\mathrm{OPE}}$ is valid.
(b) Analytic structure after the Borel transform. Since $\tilde{S}_{\mathrm{low}}$ has an infinite radius of convergence, the contour can be enlarged to $|z|>M_{\mathrm{gap}}^2$. However, $\tilde{S}_{\mathrm{OPE}}$ does not coincide with $\tilde{S}_{\mathrm{low}}$ in the shaded region even if $|z| \gg M_{\mathrm{gap}}^2$, due to duality violation.}
\label{figofconverge}
\end{figure}

Our goal is to determine the low-energy expansion coefficients $c_m$ using $S_{\mathrm{OPE}}$. The $c_m$ can be obtained by applying the Cauchy integral formula to $S_{\mathrm{low}}$ in its convergent region, but on the integration contour, the low-energy expansion cannot simply be replaced by the OPE:
\begin{align}
  \frac{c_m}{M^{2m-2j}} &= \frac{1}{2\pi i}\oint_{C_{\mathrm{small}}} dp^2 \ \frac{S_{\mathrm{low}}(p^2)}{p^{2(m+1)}}\\
  &\overset{?}{=} \frac{1}{2\pi i}\oint_{C_{\mathrm{large}}} dp^2 \ \frac{S_{\mathrm{OPE}}(p^2)}{p^{2(m+1)}}.
  \label{eq:cauchy}
\end{align}
This is because, as discussed above, the low-energy expansion converges only for $|p^2| < M_{\mathrm{gap}}^2$, whereas the OPE is valid for $|p^2| > M_{\mathrm{gap}}^2$. Thus, their regions of validity do not overlap (see figure~\ref{figofconverge}(\ref{beforeborel})).

We therefore take the Borel transform of the low-energy expansion given in eq.~\eqref{eq:LEE}: 
\begin{align}
  S_{\mathrm{low}}(p^2) = M^{2j} \sum_{m=0}^{\infty}c_m \left(\frac{p^2}{M^2}\right)^m \rightarrow \tilde{S}_{\mathrm{low}}(z) = M^{2j} \sum_{m=0}^{\infty} \frac{c_m}{m!} \left(\frac{z}{M^2}\right)^m. \label{LEEtilde}
\end{align}
After this transformation, $\tilde{S}_{\mathrm{low}}(z)$ becomes convergent at arbitrary large $|z|$. This transformation does not spoil the extraction of $c_m$ using the Cauchy integral formula. Therefore, if the transformed OPE is also valid in the region containing a contour surrounding the origin, $c_m$ can be obtained from the OPE by replacing the contour integral of $\tilde{S}_{\mathrm{low}}$ with that of $\tilde{S}_{\mathrm{OPE}}$:
\begin{align}
  \frac{c_m}{M^{2m-2j}} &= \frac{m!}{2\pi i}\oint_{C_{\mathrm{small}}} dz \ \frac{\tilde{S}_{\mathrm{low}}(z)}{z^{m+1}}\\
  &= \frac{m!}{2\pi i}\oint_{C_{\mathrm{large}}} dz \ \frac{\tilde{S}_{\mathrm{low}}(z)}{z^{m+1}}\\
  &\overset{?}{=} \frac{m!}{2\pi i}\oint_{C_{\mathrm{large}}} dz \ \frac{\tilde{S}_{\mathrm{OPE}}(z)}{z^{m+1}}.
  \label{eq:cauchyafterBorel}
\end{align}

The question is therefore whether the Borel-transformed OPE agrees with the Borel-transformed low-energy expansion along the entire contour of the Cauchy integral formula. Formally, the Borel transform of the low-energy expansion can be written as an inverse-Laplace-type integral,\footnotemark{}
\begin{align}
  \tilde{S}_{\mathrm{low}}(z) = \frac{1}{2\pi i}\int_{+0-i\infty}^{+0+i\infty}dr\, r^{-1} e^{z r} S_{\mathrm{low}}\left(p^2 = \frac{1}{r}\right), \label{eq:BorelLaplace}
\end{align}
where the same transform applied to the OPE gives
\begin{align}
  \tilde{S}_{\mathrm{OPE}}(z) = \frac{1}{2\pi i}\int_{+0-i\infty}^{+0+i\infty}dr\, r^{-1} e^{z r} S_{\mathrm{OPE}}\left(p^2 = \frac{1}{r}\right). \label{eq:dual}
\end{align}
Following ref.~\cite{Hayashi2023}, we call $\tilde{S}_{\mathrm{OPE}}$ the ``dual transform'' of $S_{\mathrm{OPE}}$.

\footnotetext{The transform, eq.~\eqref{eq:BorelLaplace}, is defined for $\operatorname{Re}(z) >0$. In the following, its values in the region $\operatorname{Re}(z) <0$ are obtained by analytic continuation.}

One might naively expect that $\tilde{S}_{\mathrm{OPE}}(z)$ inherits the region of validity $|p^2| > M_{\mathrm{gap}}^2$ of the original OPE and is therefore valid for $|z| > M_{\mathrm{gap}}^2$. This is not correct. The region in which $\tilde{S}_{\mathrm{OPE}}(z)$ provides a good approximation to the transformed exact quantity (or low-energy expansion) is restricted to $|z| > M_{\mathrm{gap}}^2$ and $\operatorname{Re}(z)>0$. Outside this region (shaded region of figure~\ref{figofconverge}(\ref{afterborel})), $\tilde{S}_{\mathrm{OPE}}(z)$ does not agree with $\tilde{S}_{\mathrm{low}}(z)$ because the integration contour passes through the low-energy region where the OPE is not valid. We explain the reason below.

\subsection{Emergence of duality violation}\label{sec:DVemargence}

Since the Borel-transformed low-energy expansion $\tilde{S}_{\mathrm{low}}(z)$ converges for arbitrary $z$, its sum coincides with the transformed exact quantity $\tilde{S}_{\mathrm{exact}}$ for any $z$ in the plane. On the other hand, the Borel-transformed OPE does not coincide with $\tilde{S}_{\mathrm{exact}}$. Denoting this difference by the ``duality-violation'' term $\tilde{S}_{\mathrm{DV}}(z)$, we have
 \begin{align}
  \tilde{S}_{\mathrm{exact}}(z) = \tilde{S}_{\mathrm{low}}(z) = \tilde{S}_{\mathrm{OPE}}(z) -\tilde{S}_{\mathrm{DV}}(z). \label{eq:OPE-DV}
 \end{align}
The OPE and the exact quantity have different analytic structures. Consequently, their Borel transforms can be represented by integrals along different contours. This difference explains the emergence of the DV term and why it becomes non-negligible for sufficiently large $|z|$ with $\operatorname{Re}(z)<0$.

\begin{figure}
\centering
\begin{minipage}[t]{0.47\linewidth}
  \centering
  \includegraphics[width=\linewidth, trim=10mm 100mm 10mm 40mm,clip]{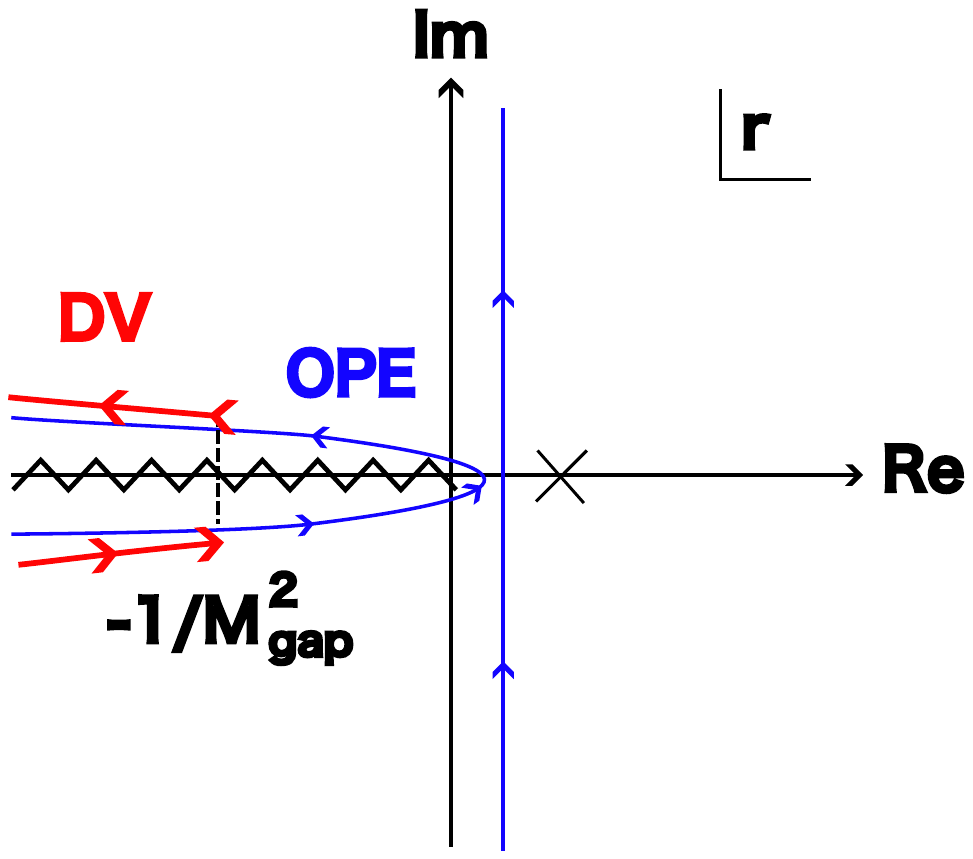}
  \par\smallskip
  \refstepcounter{jheppanel}\label{figborelOPEandDV}(\thejheppanel)
\end{minipage}
\hfill
\begin{minipage}[t]{0.47\linewidth}
  \centering
  \includegraphics[width=\linewidth, trim=10mm 100mm 10mm 40mm,clip]{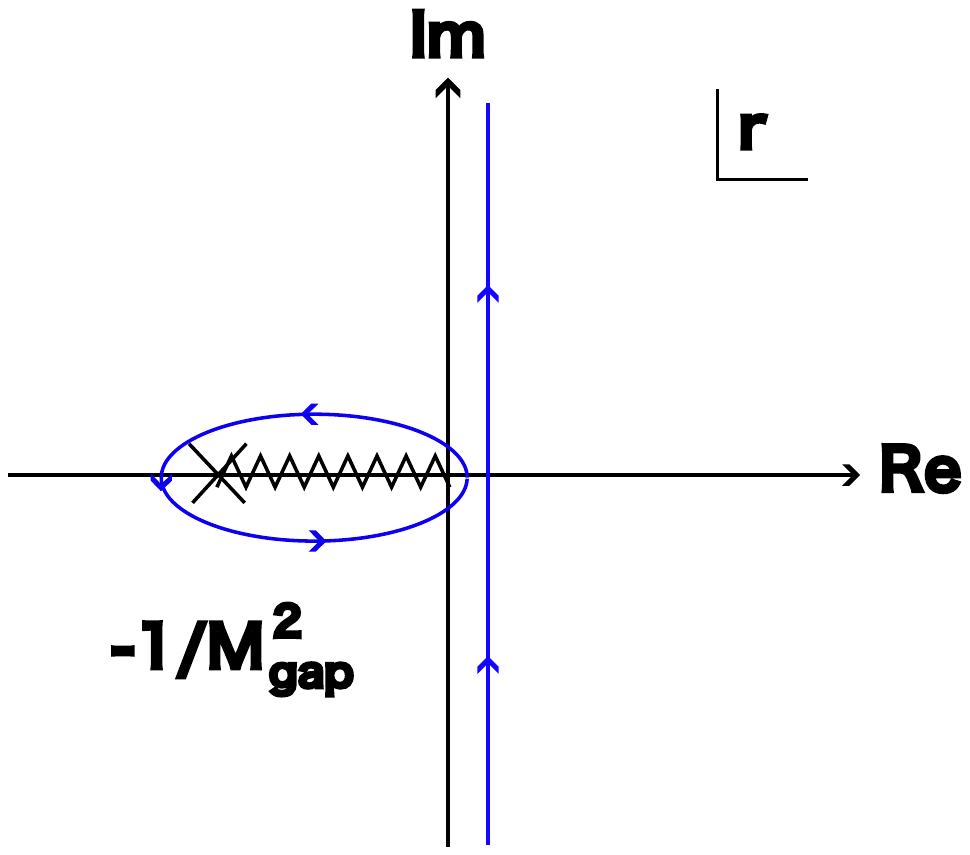}
  \par\smallskip
  \refstepcounter{jheppanel}\label{figborelLEE}(\thejheppanel)
\end{minipage}
\caption{Integration contours for the integral representation of the Borel transform.
(a) Integral contours and branch cut of $S_{\mathrm{OPE}}$. The integration contour $-i\infty+0 \to i\infty+0$ can be deformed into a contour running just above and below the branch cut along the negative real axis. The cut originates from the running coupling $g(p^2)$. The red segments (DV) denote the part of the contour for $|p^2|<M_{\mathrm{gap}}^2$, where the OPE is not valid.
(b) Integration contours and branch cut of $S_{\mathrm{exact}}$. Due to the mass gap, the branch cut consists only of the finite segment from $0$ to $-1/M_{\mathrm{gap}}^2$. Hence, the integration contour can be deformed into a closed contour surrounding the cut.}
\label{figofboreltransform}
\end{figure}

For convenience, we work in the complex $r$ plane, where $r=1/p^2$. The exact quantity $S_{\mathrm{exact}}(p^2 = 1/r)$ has a branch cut along $r \in [-1/M_{\mathrm{gap}}^2,0]$ on the real axis, and the integration contour can be wrapped around this branch cut (figure~\ref{figofboreltransform}(\ref{figborelLEE})). On the other hand, $S_{\mathrm{OPE}}(p^2=1/r)$ has a branch cut along $r \in (-\infty,0]$, originating from the analytic structure of the coupling constant $g(p^2) \sim  1/\log(p^2/M^2)$ in the Wilson coefficients (after renormalization group improvement $\mu^2 = p^2$). Thus, the integration contour of its Borel transform can be wrapped around this negative half-line (figure~\ref{figofboreltransform}(\ref{figborelOPEandDV})). Using the fact that $S_{\mathrm{OPE}}(1/r)$ approximates the exact value in $|r|<1/M_{\mathrm{gap}}^2$, the integral for the exact quantity can then be represented in terms of the OPE integral as
\begin{align}
  \tilde{S}_{\mathrm{exact}}(z) &= (\text{figure~\ref{figofboreltransform}(\ref{figborelLEE})}) \\
    &= \text{(blue segment of figure~\ref{figofboreltransform}(\ref{figborelOPEandDV}))} - \text{(red segment of figure~\ref{figofboreltransform}(\ref{figborelOPEandDV}))} \\
    &= \tilde{S}_{\mathrm{OPE}}(z) - \tilde{S}_{\mathrm{DV}}(z), \label{eq:OPE-DVfig}
\end{align}
where we have assumed that the contribution from the vicinity of the singularity at $r=-1/M_{\mathrm{gap}}^2$ is negligible.\footnotemark{} 
\footnotetext{When this contribution cannot be neglected, see the example of $\rho$ discussed below.}
Here, we define the following quantities:\footnotemark{}
\begin{align}
  \tilde{S}_{\mathrm{OPE}}(z) = \frac{1}{\pi} \int_{r=0+i0}^{-\infty+i0} dr\, r^{-1} e^{zr} \operatorname{Im}[S_{\mathrm{OPE}}(1/r)], \label{tildeSOPE}\\
  \tilde{S}_{\mathrm{DV}}(z) = \frac{1}{\pi} \int_{r=\frac{-1}{M_{\mathrm{gap}}^2}+i0}^{-\infty+i0} dr\, r^{-1} e^{zr} \operatorname{Im}[S_{\mathrm{OPE}}(1/r)].\label{tildeSDV}
\end{align}
These integral representations are defined for $\operatorname{Re}(z)>0$. In the following, their values for $\operatorname{Re}(z)<0$ are understood by analytic continuation.

Substituting $S_{\mathrm{OPE}}$ expressed in eq.~\eqref{SOPE} into eq.~\eqref{tildeSDV}, we obtain the asymptotic expansion valid for $|z|\gg M_{\mathrm{gap}}^2$,
\footnotetext{Strictly speaking, in eq.~\eqref{tildeSOPE} the divergence of the integrand near the origin must be taken into account. The divergence is canceled between the contribution from the contour encircling the origin and those from the two sides of the branch cut near the origin.}
\begin{align}
  \tilde{S}_{\mathrm{DV}}^{\mathrm{asym}}(z) \sim \sum_{l= 1}^{\infty} e^{\frac{-z}{M_{\mathrm{gap}}^2}} \left(\frac{M_{\mathrm{gap}}^2}{z}\right)^l K_l, \label{SDVasym}
\end{align}
where $K_l$ is independent of $z$ (see appendix~\ref{sec:Appdual} and next subsection).\footnotemark{} This form shows that the DV term becomes non-negligible at least when $\operatorname{Re}(z) < 0$ and $|z| \gg M_{\mathrm{gap}}^2$. For small $|z|$, the DV term can also be shown to be non-negligible by analytically evaluating the integral representation and continuing the result.
\footnotetext{In ref.~\cite{takaura2024}, this behavior was demonstrated for the explicit examples $D_\alpha$ and $\rho$ using Borel representations of their Wilson coefficients. However, it follows directly from the structure of the dual transform, without relying on the specific functional forms of those Borel representations; see appendix~\ref{sec:Appdual}.}

\subsection{Asymptotic behavior of \texorpdfstring{$\tilde{S}_{\mathrm{OPE}}$}{S OPE tilde} and the mass gap}\label{sec:StildeASYM}

After the dual transformation, $\tilde{S}_{\mathrm{OPE}}$ is in general an asymptotic series that does not converge with respect to the OPE order $n$, and therefore has an optimal truncation order $n_{\mathrm{opt}}$. This property can be used to estimate $M_{\mathrm{gap}}$. For $z>0$, evaluating \eqref{tildeSOPE} shows (see appendix~\ref{sec:Appdual} for details) that, if both $n$ and $z$ are large,
\begin{align}
  \tilde{S}_{\mathrm{OPE}}(z) \sim \sum_{n} L_{n,z} \left(\frac{M_{\mathrm{gap}}^2}{z}\right)^{n} n!, \label{OPEculc}
\end{align}
where $L_{n,z}$ does not change the leading behavior of the optimal truncation order; in particular, it contains neither factorial nor exponential dependence on $n$. The factor $(1/M_{\mathrm{gap}}^2)^{-n}$ originates from the expectation that the expansion of the OPE in $1/p^2$, centered at $r=0$, has a radius of convergence $1/M_{\mathrm{gap}}^2$. The factor $n! z^{-n}$ originates from the integral along the cut in the $r$ plane (blue segment of figure~\ref{figofboreltransform}(\ref{figborelOPEandDV})).

In general, an asymptotic series of the form
\begin{align}
  f(z) \sim \sum_{n} \left(\frac{A}{z}\right)^{n} n!\label{exampleasym}
\end{align}
is known to have an optimal truncation order $n_{\mathrm{opt}} \sim z/A$, where the absolute value of the $n$th term is minimized. In the present case, $A=M_{\mathrm{gap}}^2$. Therefore, by determining the optimal truncation order of $\tilde{S}_{\mathrm{OPE}}$ at a fixed value of $z$, an approximate value of the mass gap can be estimated from the relation $n_{\mathrm{opt}} \sim z/M_{\mathrm{gap}}^2$.

Furthermore, for an asymptotic expansion of the form in eq.~\eqref{exampleasym}, the uncertainty associated with optimal truncation is of order $e^{-z/A}$ at exponential accuracy. This is consistent with the behavior $\tilde{S}_{\mathrm{DV}}^{\mathrm{asym}} \sim e^{-z/M_{\mathrm{gap}}^2}$ in eq.~\eqref{SDVasym}.\footnotemark{}

\footnotetext{In ref.~\cite{takaura2024}, the $n$ dependence of $\tilde{S}_{\mathrm{OPE}}$ itself was not calculated directly. Instead, the asymptotic behavior of $\tilde{S}_{\mathrm{OPE}}$ was inferred by relating the behavior $e^{-z/M_{\mathrm{gap}}^2}$ of the DV term to the uncertainty of the asymptotic expansion.}

These asymptotic behaviors can be understood in a unified way from the correspondence between the integration contour of the dual transform and the integral representations of the Gamma and incomplete Gamma functions. Formally, eqs.~\eqref{tildeSOPE} and \eqref{tildeSDV} reduce to the integral representations
\begin{align}
  \Gamma(n) &= \int_{0}^{\infty} dt\ t^{n-1} e^{-t}, \\
  \Gamma(n,a) &= \int_{a}^{\infty} dt\ t^{n-1} e^{-t}.
\end{align}
Here, $n$ denotes the OPE order, while $a$ corresponds to $z/M_{\mathrm{gap}}^2$. The factorial behavior of the former corresponds to eq.~\eqref{OPEculc}, whereas the asymptotic expansion of the latter
\begin{align}
  \Gamma(n,a) \sim a^{n-1} e^{-a} \left(1 + \frac{n-1}{a} + \cdots \right)
\end{align}
corresponds to eq.~\eqref{SDVasym}. See appendix~\ref{sec:Appdual} for details.

\subsection{Extraction of \texorpdfstring{$c_0$}{c0}}
Using eq.~\eqref{eq:OPE-DVfig}, the formula \eqref{eq:cauchyafterBorel} for extracting $c_m$ becomes
\begin{align}
\frac{c_m}{M^{2m-2j}}=\frac{m!}{2\pi i}\oint \frac{dz}{z^{m+1}} \tilde{S}_{\mathrm{low}}(z) = \frac{m!}{2\pi i}\oint \frac{dz}{z^{m+1}} (\tilde{S}_{\mathrm{OPE}}(z) -\tilde{S}_{\mathrm{DV}}(z)).\label{eq:extract}
\end{align}
Thus, the term $\tilde{S}_{\mathrm{DV}}$ gives an additional (unwanted) contribution when extracting $c_m$ from $\tilde{S}_{\mathrm{OPE}}$. Hence, ref.~\cite{takaura2024} exploits the fact that $\tilde{S}_{\mathrm{OPE}}(z)$ provides a good approximation in the right-half plane, whereas $\tilde{S}_{\mathrm{DV}}(z)$ becomes non-negligible in the left-half plane. By multiplying $\tilde{S}_{\mathrm{OPE}}$ by a weight function $F(z)$ that is large in the right-half plane and small in the left-half plane, the DV contribution can be suppressed and $c_m$ can be extracted from $\tilde{S}_{\mathrm{OPE}}$ alone. A specific weight function for $m=0$ was constructed as 
\begin{align}
  \frac{c_{0}}{M^{-2j}} &\approx \frac{1}{2\pi i}\oint_{z=z_0 e^{i\theta}} \frac{dz}{z} F_{(0)}(z) \tilde{S}_{\mathrm{OPE}}(z),\\
  F_{(0)}(z) &= \left(\frac{z}{z_0} +1\right)^k, \qquad 1 \ll k=\frac{z_0}{M_{\mathrm{gap}}^2} \in \mathbb{N}. \label{eq:F0tak}
\end{align}
Indeed, since $F_{(0)}(0)=1$ and $F_{(0)}(z)$ contains only nonnegative powers of $z$, it leaves the $z^0$ coefficient of $\tilde{S}_{\mathrm{low}}$ unchanged. At the same time, choosing $z_0$ such that $k\gg1$ makes $F_{(0)}$ small in the left-half plane, suppressing the DV contribution to $O(M_{\mathrm{gap}}^2/z_0)$.

Furthermore, by taking $z_0$ sufficiently large in numerical calculations, the remaining $z_0$-dependent error can be made arbitrarily small. The dual-transformed OPE $\tilde{S}_{\mathrm{OPE}}$ also has an optimal truncation order $n_{\mathrm{opt}}\sim z_0/M_{\mathrm{gap}}^2$, with an associated truncation uncertainty. Since this order increases with $z_0$, the accuracy of the calculation can in principle be improved arbitrarily if the OPE is known to sufficiently high orders.

This weight function $F_{(0)}$ is specific to the extraction of $c_0$ (or the first nonzero coefficient). For higher-order coefficients, it would generally mix different terms of the low-energy expansion. This motivates the generalized weight function introduced in the next section.

\section{Improved methods}\label{sec:improve}
The method of ref.~\cite{takaura2024} described above provides a new way to determine low-energy quantities from the OPE, but it was applied only to the lowest-order coefficient $c_0$ of the low-energy expansion. We propose two extensions of this method that allow the low-energy expansion coefficients $c_m$ at arbitrary orders to be calculated.

Method I generalizes the weight function of the previous method. Method II instead directly constructs $\tilde{S}_{\mathrm{OPE}}-\tilde{S}_{\mathrm{DV}}$.

\subsection{Method I: general form of weight function \texorpdfstring{$F_m(z)$}{Fm(z)}}\label{sec:methodI}
In Method I, we construct a generalized weight function $F_m(z)$ that can be used to determine $c_m$. We replace the qualitative requirement that the weight function be ``small in the region where the DV term is large'' with explicit conditions on $F_m(z)$.

We require $F_m(z)$ to satisfy two conditions: it does not interfere with the extraction of $c_m$ from the low-energy expansion; it eliminates the contribution from the DV term. These conditions can be written, respectively, as 
\begin{align}
  \frac{m!}{2\pi i}\oint \frac{dz}{z^{m+1}} F_m(z) \tilde{S}_{\mathrm{low}}(z) &= \frac{c_m}{M^{2m-2j}}, \label{eq:condLEE}\\
  \frac{m!}{2\pi i}\oint \frac{dz}{z^{m+1}} F_m(z) \tilde{S}_{\mathrm{DV}}(z) &= 0. \label{eq:condDV}
\end{align}
If a function satisfying both conditions is obtained, the low-energy expansion coefficients can be extracted by
\begin{align}
  \frac{m!}{2\pi i}\oint \frac{dz}{z^{m+1}} F_m(z) \tilde{S}_{\mathrm{OPE}}(z) =  \frac{c_m}{M^{2m-2j}}.
\end{align}
In the following, we determine a function $F_m(z)$ that satisfies \eqref{eq:condLEE} and \eqref{eq:condDV} for general $m$.

First, we assume that $F_m(z)$ admits a Laurent expansion around the origin. This allows us to apply the residue theorem.
\begin{align}
  F_m(z) = \sum_{s=-s_0}^{\infty} f_s^{(m)} \left(\frac{z}{z_0}\right)^{s},\label{eq:defF}
\end{align}
where $s_0$ is an arbitrary nonnegative integer and $f_s^{(m)}$ is a constant independent of $z$.

Substituting $\tilde{S}_{\mathrm{low}}$ in eq.~\eqref{LEEtilde} into the condition \eqref{eq:condLEE} and applying the residue theorem, we obtain
\begin{align}
  \frac{c_m}{M^{2m-2j}} = m! \sum_{s= -s_0}^{m} f_s^{(m)} \left(\frac{1}{z_0}\right)^s \frac{c_{m-s}}{(m-s)!} \frac{1}{M^{2(m-s-j)}}.
\end{align}
Consistency between both sides requires $f_0^{(m)} = 1$ and $f_s^{(m)} = 0$ for $-s_0 \le s \le -1$ and $1\le s\le m$. Since the low-energy expansion contains only nonnegative powers of $p^2$, the coefficients $f_s^{(m)}$ for $s\ge m+1$ do not contribute to the residue and remain arbitrary. Therefore, a weight function satisfying the condition \eqref{eq:condLEE} must have the form
\begin{align}
  F_m(z) = 1+ \sum_{s= m+1}^{\infty} f_{s}^{(m)} \left(\frac{z}{z_0}\right)^{s}.\label{eq:FfromLEE}
\end{align}

Next, we consider the condition \eqref{eq:condDV}, which requires the contribution from the DV term to vanish. Unlike the condition \eqref{eq:condLEE}, it is difficult to satisfy this condition exactly.\footnotemark{} We therefore seek a weight function that eliminates the contribution from the asymptotic expansion \eqref{SDVasym} of $\tilde{S}_{\mathrm{DV}}$.

\footnotetext{The exact form of $\tilde{S}_{\mathrm{DV}}(z)$ has a branch point at the origin, so that the residue theorem cannot be applied.}

Substituting the asymptotic expansion into eq.~\eqref{eq:condDV} and requiring $F_m(z)$ to be independent of the coefficients $K_l$, which depend on the specific coefficients of the OPE, we obtain
\begin{align}
    0 = \frac{m!}{2\pi i}\oint dz\ \frac{1}{z^{m+1}} F_m(z) e^{-\frac{z}{M_{\mathrm{gap}}^2}} \left(\frac{M_{\mathrm{gap}}^2}{z}\right)^{l} \label{eq:FcondDVint}
\end{align}
for each $l=1,2,\cdots$. If the integrand contains no $z^{-1}$ term, this condition is satisfied. Since eq.~\eqref{eq:FcondDVint} must hold for every $l\ge1$, all coefficients of $z^{m+l}$ in $e^{-z/M_{\mathrm{gap}}^2}F_m(z)$ must vanish. Therefore, $F_m(z)$ must be of the form
\begin{align}
  F_m(z) = e^{\frac{z}{M_{\mathrm{gap}}^2}} \times (\text{polynomial in $z$ of degree at most $m$}). \label{eq:Famakudari}
\end{align}
Equivalently, $e^{-z/M_{\mathrm{gap}}^2}F_m(z)$ is a polynomial of degree at most $m$, so each term in the asymptotic expansion of the DV contribution satisfies the condition \eqref{eq:FcondDVint}.

We have thus obtained the requirements \eqref{eq:FfromLEE} and \eqref{eq:Famakudari}. The two requirements uniquely determine the weight function
\begin{align}
  F_m(z) &= e^{\frac{z}{M_{\mathrm{gap}}^2}} \sum_{s=0}^{m}\frac{1}{s!} \left(\frac{-z}{M_{\mathrm{gap}}^2}\right)^s =\frac{\Gamma\left(m+1, -\frac{z}{M_{\mathrm{gap}}^2}\right)}{\Gamma(m+1)}.\label{eq:F}
\end{align}

We therefore find that low-energy expansion coefficients at arbitrary orders can be extracted from the OPE through
\begin{align}
  \frac{m!}{2\pi i}\oint_{z=z_0 e^{i\theta}} \frac{dz}{z^{m+1}} \frac{\Gamma\left(m+1, -\frac{z}{M_{\mathrm{gap}}^2}\right)}{\Gamma(m+1)} \tilde{S}_{\mathrm{OPE}}(z) \simeq \frac{c_m}{M^{2m-2j}}. \label{eq:result}
\end{align}
Here, we use ``$\simeq$'' rather than ``$=$'' because the weight function satisfies only the condition for the asymptotic expansion of the DV term. The discrepancy between the two sides therefore originates from the part of the DV term that cannot be represented by its asymptotic expansion. In particular, this approximation is not justified for $z_0\lesssim M_{\mathrm{gap}}^2$, where the asymptotic expansion is not applicable. The approximation improves as $z_0$ becomes large.

An additional error arises from truncating the OPE at finite order $n$. As in the method of ref.~\cite{takaura2024}, this error can in principle be reduced by increasing the OPE truncation order $n$ and $z_0$ simultaneously, keeping $n$ near $n_{\mathrm{opt}}\sim z_0/M_{\mathrm{gap}}^2$.

\subsection{Weight function \texorpdfstring{$F_{(0)}$}{F(0)} in previous work}\label{sec:c0hikaku}
The weight function $F_{(0)}$ proposed in ref.~\cite{takaura2024} can be characterized in terms of the conditions imposed in the previous subsection. In fact, $F_{(0)}$ partially satisfies these conditions.

Since
\begin{align}
  \left(\frac{z}{z_0}+1\right)^{\frac{z_0}{M_{\mathrm{gap}}^2}} = 1 + \frac{z_0}{M_{\mathrm{gap}}^2} \frac{z}{z_0} + \dots,
\end{align}
it exactly satisfies eq.~\eqref{eq:FfromLEE} for $m=0$. On the other hand, substituting $F_{(0)}$ into eq.~\eqref{eq:FcondDVint}, with $k\ge l$, gives
\begin{align}
\frac{1}{2\pi i} \oint \frac{dz}{z} F_{(0)}(z) e^{\frac{-z}{M_{\mathrm{gap}}^2}}\left(\frac{M_{\mathrm{gap}}^2}{z}\right)^l &= \sum_{r=0}^{l}\binom{k}{l-r}\frac{(-k)^r}{r!} k^{-l},
\end{align}
where $k=z_0/M_{\mathrm{gap}}^2 \in \mathbb{N}$. For $l=1$, this term vanishes. For $l\ge 2$, however, contributions of order $O(M_{\mathrm{gap}}^2/z_0)$ remain.\footnotemark{} By contrast, the weight function \eqref{eq:F} makes the contributions vanish for arbitrary $l$.

\footnotetext{The weight function $F_{(0)}$ can also be generalized to extract higher-order coefficients, retaining the same
$O(M_{\mathrm{gap}}^2/z_0)$ suppression of DV contributions. We do not pursue this generalization in this paper.}

\subsection{Method II: subtraction of DV part from OPE}\label{sec:methodII}

In Method II, we directly construct $\tilde{S}_{\mathrm{OPE}}-\tilde{S}_{\mathrm{DV}}$ and extract $c_m$, rather than using a weight function as in Method I.

Indeed, the procedure for Method I,
\begin{enumerate}
  \item Calculate $\tilde{S}_{\mathrm{OPE}}$ from $S_{\mathrm{OPE}}$.
  \item Determine $M_{\mathrm{gap}}$ from $\tilde{S}_{\mathrm{OPE}}$.
  \item Construct the weight function $F_m$ (using $M_{\mathrm{gap}}$).
  \item Extract $c_m$ by the weighted contour integral in $z$,
\end{enumerate}
can be replaced by 
\begin{enumerate}
  \item Calculate $\tilde{S}_{\mathrm{OPE}}$ from $S_{\mathrm{OPE}}$.
  \item Determine $M_{\mathrm{gap}}$ from $\tilde{S}_{\mathrm{OPE}}$.
  \item Calculate $\tilde{S}_{\mathrm{OPE}}-\tilde{S}_{\mathrm{DV}}$ (using $M_{\mathrm{gap}}$ and $S_{\mathrm{OPE}}$).
  \item Extract $c_m$ from the $z$-expansion of $\tilde{S}_{\mathrm{OPE}}-\tilde{S}_{\mathrm{DV}}$,
\end{enumerate}
to implement Method II.\footnotemark{}

\footnotetext{In ref.~\cite{takaura2024}, $c_m$ was not extracted using this method. However, in the calculation presented in the appendix of that work, it was pointed out that the combination of the OPE and the DV term is a meaningful quantity when relating the OPE to the exact physical quantity.}

In particular, $\tilde{S}_{\mathrm{OPE}}-\tilde{S}_{\mathrm{DV}}$ corresponds to the blue segment minus the red segment in figure~\ref{figofboreltransform}(\ref{figborelOPEandDV}), and can therefore be obtained directly by restricting the integration contour to $r=0 \to -1/M_{\mathrm{gap}}^2$. Denoting this contour by $C_{\mathrm{OPE-DV}}$, we define
\begin{align}
  \tilde{S}_{\mathrm{OPE-DV}}(z) = \frac{1}{2\pi i} \int_{C_{\mathrm{OPE-DV}}} dr\, r^{-1} e^{zr} S_{\mathrm{OPE}}(1/r). \label{StildeOPE-DV}
\end{align}
The integrand diverges near the origin, but this divergence is canceled between the contribution from a small contour around the origin and the contributions from the two sides of the cut near the origin. Therefore, one should not simply integrate the discontinuity over $-1/M_{\mathrm{gap}}^2<r<0$.

In fact, this integral can be written in the form
\begin{align}
    \tilde{S}_{\mathrm{OPE-DV}}(z) = \sum_{m=0}^{\infty} J_m \frac{z^m}{m!}. \label{OPE-DVculc}
\end{align}
Here, $J_m$ is independent of $z$ and is determined by $S_{\mathrm{OPE}}$ and $M_{\mathrm{gap}}$. Its explicit expression, involving sums over $n$ and $k$, is given by eq.~\eqref{eq:explicitJm} in appendix~\ref{sec:AppDerivOPE-DV}. 

Equation~\eqref{OPE-DVculc} has the same power-series structure as the Borel-transformed low-energy expansion \eqref{LEEtilde}. Since $\tilde{S}_{\mathrm{OPE-DV}}$ corresponds to $\tilde{S}_{\mathrm{low}}$, we can identify their coefficients and obtain
\begin{align}
  \frac{c_m}{M^{2m-2j}}= J_m \label{cm=Jm}
\end{align}
without performing the contour integral over $z$. Since neither the $z$-integral nor the large-$z$ asymptotic expansion of the DV term is used, the result is independent of $z_0$.

As in section~\ref{sec:StildeASYM}, the power-series form of $\tilde{S}_{\mathrm{OPE-DV}}$ can be understood from the lower incomplete Gamma function,
\begin{align}
  \gamma(n,a) = \int_{0}^{a} dt\, t^{n-1}e^{-t} = \sum_{m=0}^{\infty} \frac{(-1)^m a^{n+m}}{m! (n+m)}.
\end{align}
Recalling that $a$ corresponds to $z/M_{\mathrm{gap}}^2$ in the present case, one finds that the series on the right-hand side reduces to the form of eq.~\eqref{OPE-DVculc}.

\subsection{Convergence and accuracy}\label{sec:convergence}

In the process of extracting the low-energy expansion coefficients $c_m$ from the OPE, two types of divergence can arise. The first is the divergence already present in the OPE itself, namely, the factorial divergence of the perturbative expansion of the Wilson coefficients given in eq.~\eqref{eq:WilsonPT}, $d_{n,k} g^{k+1} \sim k! g^{k+1}$. However, it is known that this renormalon-type factorial divergence is suppressed by the dual transform defined in eq.~\eqref{eq:dual}~\cite{Hayashi2023}, which is introduced when applying the Borel transform of the low-energy expansion to the OPE.\footnotemark{}

\footnotetext{More precisely, it has been shown that the factorial divergence is suppressed. Other divergences weaker than factorial growth may remain, and convergence of the perturbative expansion after the transformation is therefore not guaranteed.}

The second is a divergence with respect to the OPE order $n$ which arises from the dual transform from $S_{\mathrm{OPE}}$ to $\tilde{S}_{\mathrm{OPE}}$. As seen in eq.~\eqref{OPEculc}, the transformed OPE has a factorial $n!$ growth and is therefore an asymptotic series in $n$. Consequently, the extraction formula eq.~\eqref{eq:result} of Method I has an optimal truncation order $n_{\mathrm{opt}} \sim z_0/M_{\mathrm{gap}}^2$ and an associated uncertainty. Therefore, to improve the accuracy, both $z_0$ and the truncation order must be increased simultaneously, as in the method of ref.~\cite{takaura2024}. 

By contrast, in Method II, the coefficients $J_m$ in eq.~\eqref{OPE-DVculc} are given by sums over the OPE order $n$ without the additional $n!$-type growth introduced by the dual transform. Hence, there is no corresponding optimal-truncation limitation, and the accuracy can be improved simply by including higher-order terms in the OPE.

For Method I, as well as the method of ref.~\cite{takaura2024}, errors arise not only from truncating the OPE but also from the finite contour radius $z_0$. Since the weight function $F_m(z)$ eliminates only the contributions from the large-$z$ asymptotic expansion $\tilde{S}_{\mathrm{DV}}^{\mathrm{asym}}$, the remainder of the DV term still contributes to the error. This error decreases as $z_0$ increases but may be amplified at higher $m$.

The difference between Methods I and II in convergence properties and numerical accuracy will also be examined in the explicit examples in the next section. 

\section{Examples in the large-\texorpdfstring{$N$}{N} \texorpdfstring{$O(N)$}{O(N)} nonlinear sigma model}

In this section, we verify through explicit calculations that the low-energy expansion coefficients $c_m$ can be extracted from the OPE using the two methods proposed in section~\ref{sec:improve}. We also compare their convergence properties with respect to the OPE truncation order $n$ and their numerical accuracy, as discussed in section~\ref{sec:convergence}.

As a toy model in which the low-energy expansion can be calculated exactly, we use the large-$N$ limit of the two-dimensional $O(N)$ nonlinear sigma model, as in ref.~\cite{takaura2024}. The action is given by
\begin{align}
  S = \frac{1}{2 g_0} \int d^2x \{\partial_{\mu} \sigma^a(x)\partial_{\mu} \sigma^a(x) +\alpha(x)(\sigma^a(x)\sigma^a(x) -N)\}, \quad (a=1,2,\dots,N).
\end{align} 
In the large-$N$ limit, the running coupling constant is exactly given by
\begin{align}
  g(\mu^2) = \frac{1}{\log(\frac{\mu^2}{M^2})},
\end{align}
where $M$ is the dynamically generated mass scale.

We consider the following two quantities:
\begin{align}
  D_{\alpha}(p^2) &= N\int d^2x\ e^{-ip(x-y)}\langle\delta\alpha(x)\delta\alpha(y) \rangle, \label{eq:dalpha}\\
  \rho(p^2) &= p^2 \left(\frac{d}{d(p^2)}\right)^2 \Pi_{\sigma}(p^2),\label{eq:rho}\\
  \Pi_{\sigma}(p^2) &= \int\frac{d^2k}{(2\pi)^2}\frac{1}{(p+k)^2+M^2}D_{\alpha}(k^2).\label{eq:rhoPi}
\end{align}
$D_{\alpha}$ represents the propagator of the fluctuation $\delta \alpha = \alpha - \langle \alpha \rangle$ of the auxiliary field $\alpha(x)$ around its vacuum expectation value $\langle \alpha \rangle = M^2 $. The quantity $\rho$ is obtained by applying derivatives to the $\sigma$-field self-energy correction $\Pi_{\sigma}$ at order $1/N$ to make it UV finite.

$D_{\alpha}$ has no IR renormalon divergence, whereas $\rho$ has IR renormalons~\cite{Hayashi2023,Beneke1999}. The threshold of $D_{\alpha}$ is $M_{\mathrm{gap}}^2 = 4M^2$, while that of $\rho$ is $M_{\mathrm{gap}}^2 = 9M^2$. To focus on the extraction of the low-energy coefficients, we use these exact values of $M_{\mathrm{gap}}^2$ in the numerical calculations below and later examine the dependence of $c_m$ on $M_{\mathrm{gap}}^2$. For simplicity, we set $M^2=1$.

\subsection{Propagator of auxiliary-field fluctuation \texorpdfstring{$\delta \alpha$}{delta alpha}: \texorpdfstring{$D_\alpha$}{D alpha}}
Since $D_{\alpha}$ has mass dimension 2, it corresponds to $j=1$ in eq.~\eqref{eq:OPE}. In the large-$N$ limit, it is exactly given by~\cite{NOVIKOV1984103}
\begin{align}
  D_\alpha(p^2) = \frac{4\pi \sqrt{p^2 (p^2+4M^2)}}{\log \left(\frac{\sqrt{p^2+4M^2}+\sqrt{p^2}}{\sqrt{p^2+4M^2}-\sqrt{p^2}}\right)}.
\end{align}
Both its OPE and low-energy expansion can also be calculated exactly~\cite{takaura2024},
\begin{align}
  &D_{\alpha,\mathrm{OPE}} = p^2 \sum_{n=0}^{\infty}C_n(p^2)\left(\frac{M^2}{p^2}\right)^n, \\
  &C_0(p^2) = 4\pi g(p^2),\quad C_1(p^2)=4\pi(2g(p^2)-2g(p^2)^2),\quad \dots.
\end{align}
\begin{align}
  &D_{\alpha,\mathrm{low}} =M^2 \sum_{m=0}^{\infty}c_m \left(\frac{p^2}{M^2}\right)^m,\\
  &c_0=8\pi,\quad c_1=4\pi/3,\quad \dots.
\end{align}
For each $n$, the Wilson coefficient $C_n$ contains only finitely many powers of $g(p^2)$, and hence its perturbative expansion has no renormalon divergence.

In ref.~\cite{takaura2024}, the dual transform of $D_{\alpha}$ was calculated through a Borel transform with respect to the coupling-constant expansion. Here, instead, we perform the calculation explicitly in terms of $d_{n,k}$. This allows us to examine more directly the origin of the $n!$-type behavior of the dual-transformed OPE and the $e^{-z/M_{\mathrm{gap}}^2}$-type behavior of the DV term. 

\begin{figure}[tb]
  \centering
  \begin{minipage}[t]{0.49\textwidth}
    \centering
    \includegraphics[width=\linewidth]{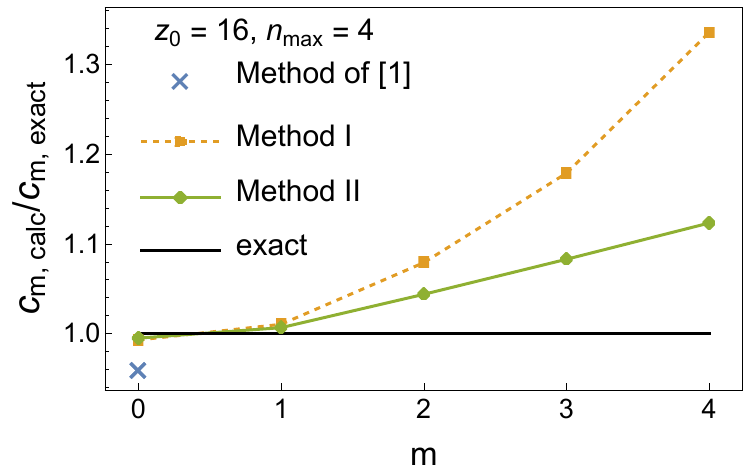}
    \par\smallskip
    \refstepcounter{jheppanel}\label{fig:dalpha-m}(\thejheppanel)
  \end{minipage}
  \hfill
  \begin{minipage}[t]{0.49\textwidth}
    \centering
    \includegraphics[width=\linewidth]{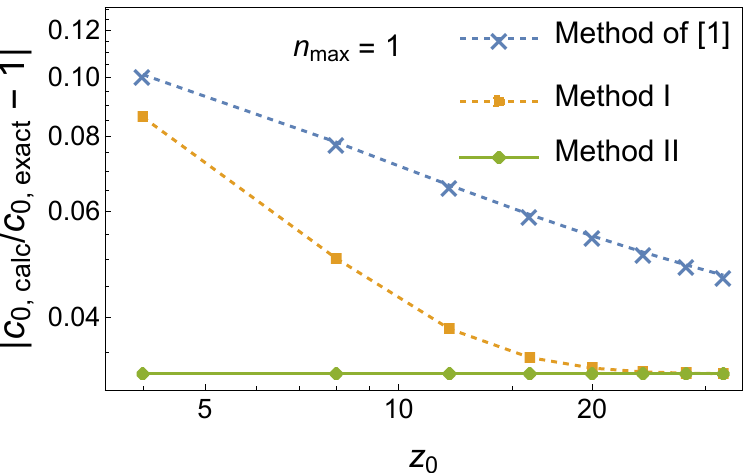}
    \par\smallskip
    \refstepcounter{jheppanel}\label{fig:dalpha-z0}(\thejheppanel)
  \end{minipage}
  \caption{Application of the methods to $D_{\alpha}$: higher-order coefficients $c_m$ and $z_0$-dependence.
  (a) $c_m$ of $D_{\alpha}$ calculated using Methods I and II and the method of ref.~\cite{takaura2024} (only for $c_0$), compared with the exact values. The vertical axis shows $c_{m,\mathrm{calc}}/c_{m,\mathrm{exact}}$. $n_{\mathrm{max}}=4$ and $z_0=16$ for Method I and the method of ref.~\cite{takaura2024}.
  (b) Dependence of the relative error in $c_0$ of $D_{\alpha}$ on $z_0$, comparing Method I and the method of ref.~\cite{takaura2024} with the $z_0$-independent result of Method II. $n_{\mathrm{max}}=1$. Both axes are logarithmic.}
  \label{fig:dalpha-results1}
\end{figure}

Figure~\ref{fig:dalpha-results1}(\ref{fig:dalpha-m}) shows the ratios of the calculated low-energy expansion coefficients $c_m$ to their exact values for $m=0,1,2,3,4$. The results are obtained using Method I, eq.~\eqref{eq:result}, and Method II, eq.~\eqref{cm=Jm}. For comparison, we also show $c_0$ obtained using the method of ref.~\cite{takaura2024} (eq.~\eqref{eq:F0tak}). 

Both Methods I and II extract higher-order $c_m$ with reasonable accuracy. As $m$ increases, the accuracy of both Methods I and II becomes worse due to the OPE truncation error at fixed $n_{\mathrm{max}}$. The difference between Methods I and II also increases with $m$, reflecting the amplification of the $z_0$-dependent error in Method I. Thus, Method II is more accurate over the range shown.

The calculations involve the OPE truncation order $n_{\mathrm{max}}$ and coupling-constant expansion truncation order $k_{\mathrm{max}}$. For $D_{\alpha}$, however, the coupling-constant expansion terminates at a finite order for each $n$, so all perturbative terms up to the chosen $n_{\mathrm{max}}$ can be included. For Method I and the method of ref.~\cite{takaura2024}, the contour radius $z_0$ is also required.

In figure~\ref{fig:dalpha-results1}(\ref{fig:dalpha-m}), we set $z_0=16$ and $n_{\mathrm{max}} = 4$, where $n_{\mathrm{max}}$ is equal to the estimated optimal truncation order $z_0/M_{\mathrm{gap}}^2$. Since $z_0/M_{\mathrm{gap}}^2=4$, the condition $z_0/M_{\mathrm{gap}}^2 \gg 1$ assumed in the asymptotic expansion \eqref{SDVasym} (and, for the method of ref.~\cite{takaura2024}, in eq.~\eqref{eq:F0tak}) is not well satisfied. Nevertheless, the low-order coefficients $c_m$ ($m=0,1,2$) can be determined with errors of about 10\%.

Figure~\ref{fig:dalpha-results1}(\ref{fig:dalpha-z0}) shows the $z_0$ dependence of the accuracy in the calculation of $c_0$ on logarithmic scales for both axes.\footnotemark{} The vertical axis represents the magnitude of the relative error, hence the values closer to zero indicate higher accuracy. For comparison, we also show the result obtained with Method II, which is independent of $z_0$. The remaining error in Method II can be attributed to truncating the OPE at $n_{\mathrm{max}} =1$. For both the method of ref.~\cite{takaura2024} and Method I, the $z_0$-dependent error decreases as $z_0$ increases, and their results approach that of Method II, for which only the OPE truncation error remains. The $z_0$-dependent error is $O(z_0^{-1})$ in the method of ref.~\cite{takaura2024}, whereas it decreases more rapidly in Method I. This is consistent with section~\ref{sec:c0hikaku}, where we found that Method I imposes a stronger suppression of the DV contribution.

\footnotetext{We show the result for $c_0$ in order to compare our methods with the method of ref.~\cite{takaura2024}. Similar behavior is also observed for $m>0$ in our methods. The same applies to the other results.}

\begin{figure}[tb]
  \centering
  \begin{minipage}[t]{0.49\textwidth}
    \centering
    \includegraphics[width=\linewidth]{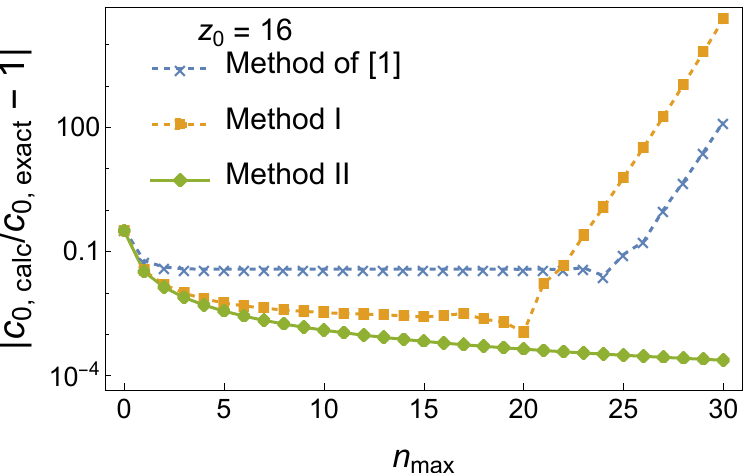}
    \par\smallskip
    \refstepcounter{jheppanel}\label{fig:dalpha-nmax}(\thejheppanel)
  \end{minipage}
  \hfill
  \begin{minipage}[t]{0.49\textwidth}
    \centering
    \includegraphics[width=\linewidth]{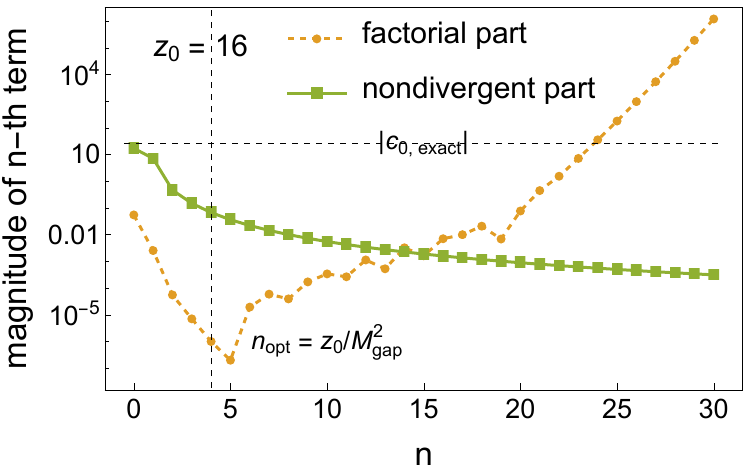}
    \par\smallskip
    \refstepcounter{jheppanel}\label{fig:dalpha-nmax-fact}(\thejheppanel)
  \end{minipage}
  \caption{Application of the methods to $D_{\alpha}$: dependence on the OPE truncation order.
  (a) Dependence of the relative error in $c_0$ of $D_{\alpha}$ on the OPE truncation order $n_{\mathrm{max}}$, comparing the convergence of the three methods. $z_0=16$ for Method I and the method of ref.~\cite{takaura2024}. The vertical axis is logarithmic.
  (b) Magnitudes of the $n$th-order contributions to $c_0$ on a logarithmic scale. The orange line shows the difference between the contributions from Methods I and II, while the green line shows the Method II contribution. The reference lines indicate $n=z_0/M_{\mathrm{gap}}^2=4$ and $|c_{0,\mathrm{exact}}|$.}
  \label{fig:dalpha-results2}
\end{figure}

Figure~\ref{fig:dalpha-results2}(\ref{fig:dalpha-nmax}) shows the dependence on the OPE truncation order $n_{\mathrm{max}}$. The vertical axis, shown on a logarithmic scale, again represents the magnitude of the relative error. The error increases rapidly beyond $n_{\mathrm{max}} \sim 20$ for Method I and beyond $n_{\mathrm{max}} \sim 25$ for the method of ref.~\cite{takaura2024}. In contrast, the error in Method II continues to decrease throughout the range shown. At first sight, this may appear inconsistent with the statement that $\tilde{S}_{\mathrm{OPE}}$ has an optimal truncation order at $n \sim z/M_{\mathrm{gap}}^2 =4$. However, this apparent discrepancy does not invalidate the estimate given in eq.~\eqref{OPEculc}, as explained below.

To understand this behavior, see figure~\ref{fig:dalpha-results2}(\ref{fig:dalpha-nmax-fact}). The horizontal axis is the OPE order $n$, and the vertical axis, shown on a logarithmic scale, is the magnitude of the contribution from the term at each order, rather than the cumulative sum up to $n_{\mathrm{max}}$. The orange line shows the contribution of the $n$th-order term in Method I, eq.~\eqref{eq:result}, after subtracting the corresponding Method II contribution, which does not contain the $n!$-type divergence. This quantity reaches its minimum around $n\sim5$,\footnotemark{} which is consistent with the estimate $z_0/M_{\mathrm{gap}}^2 =4$. On the other hand, the green line shows the contribution of Method II at each order. The $n=0$ term in Method II accounts for most of the exact value. Although the factorially growing component (orange line) starts increasing around $n\sim 5$, its magnitude remains smaller than that of the $n=0$ term in Method II until around $n\sim 25$.
\footnotetext{A deviation of order $O(1)$ can naturally arise from subleading corrections.}

Thus, although the factorially growing component has already begun to increase beyond its optimal truncation order, it initially remains much smaller than the nondivergent contribution from low orders in $n$. As a result, the order at which the divergence becomes visible in the final result can be substantially larger than the optimal truncation order estimated from eq.~\eqref{OPEculc}.

The same mechanism also applies to the method of ref.~\cite{takaura2024}, with the onset of the visible divergence occurring at a slightly higher order.

In the present toy model, the exact mass gap $M_{\mathrm{gap}}^2=4$ is known. In general, however, it is difficult to determine the mass gap exactly, and even when it can be estimated, some uncertainty is typically involved. We therefore examine the sensitivity of the results by assuming different values of $M_{\mathrm{gap}}^2$, as shown in table~\ref{tab:dalpha-massgap}.
We set $n_{\mathrm{max}}=4$ and, for Method I, $z_0=16$. The table shows the relative errors with respect to the exact value $c_0=8\pi$. Both methods are fairly insensitive to variations of about $10\%$ in the assumed value of $M_{\mathrm{gap}}^2$. Even when $M_{\mathrm{gap}}^2$ is doubled or halved, the relative error in the extracted $c_0$ remains below 10\%.

\begin{table}
  \centering
  \begin{tabular}{c|ccccc}
    \hline
    $M_{\mathrm{gap}}^2$ & $2$ & $3.6$ & $4\,\text{(exact)}$ & $4.4$ & $8$ \\
    \hline
    Method I  & $+2.6\%$ & $-1.1\%$ & $-0.71\%$ & $-0.17\%$ & $+6.1\%$ \\
    Method II & $+2.6\%$ & $-0.97\%$ & $-0.50\%$ & $+0.15\%$ & $+9.1\%$ \\
    \hline
  \end{tabular}
  \caption{Relative errors in the extracted $c_0$ of $D_{\alpha}$ for different input values of $M_{\mathrm{gap}}^2$. $z_0=16$ and $n_{\mathrm{max}}=4$.}
  \label{tab:dalpha-massgap}
\end{table}

\subsection{Self-energy correction of \texorpdfstring{$\sigma$}{sigma}: \texorpdfstring{$\rho$}{rho}}

Since $\rho$ has mass dimension $0$, it corresponds to $j=0$ in eq.~\eqref{SOPE}. The low-energy expansion can be calculated numerically from the definition in eqs.~\eqref{eq:rho} and \eqref{eq:rhoPi}~\cite{takaura2024}:
\begin{align}
  &\rho_{\mathrm{low}}(p^2) =\sum_{m=0}^{\infty}c_m \left(\frac{p^2}{M^2}\right)^m,\\
  & c_0=0,\quad c_1=-0.012396\dots,\quad c_2=0.0014307\dots,\quad \dots.
\end{align}

On the other hand, the calculation of the OPE is more involved. Therefore, as in ref.~\cite{takaura2024}, we use the Borel representation of the OPE with respect to the coupling-constant expansion. We refer to ref.~\cite{Hayashi2023} for the details and present only the results used here. For $\rho$, the mass gap is $M_{\mathrm{gap}}=3M$, corresponding to the sum of the $\sigma$-field mass $M$ and the $\delta\alpha$ threshold $2M$.

By substituting the OPE of $D_{\alpha}$ into the definition of $\rho$, one obtains the Borel-transformed OPE:\footnotemark{}
\begin{align}
  B_{\rho,\mathrm{OPE}}(p^2, u) &=\frac{1}{4\pi}\left(\frac{\mu^2}{p^2}\right)^u \sum_{n=0}^{\infty}\left(\frac{M^2}{p^2}\right)^n \{a_n(u)\log(p^2/M^2)+b_n(u) \},\\
  a_n(u) &= \sum_{j=0}^{n}\frac{(-1)^{n-j}}{(n-j)!^2}X_j(u) ((j+u-1)_{n-j})^2 (n+u)(n+u-1),\\
  b_n(u)&=\overline{b}_{n}(u)-\pi\frac{\cos(\pi u)}{\sin(\pi u)}a_n(u),\\
  \overline{b}_{n}(u)&=\sum_{j=0}^{n}\frac{(-1)^{n-j}}{(n-j)!^2}X_j(u)((j+u-1)_{n-j})^2 \notag\\
  &\qquad \{1-2(n+u)+2(n+u)(n+u-1)(\psi(1+n-j)-\psi(n+u-1))\}.
\end{align}
Here, we use the Pochhammer symbol $(X)_Y = \Gamma(X+Y)/\Gamma(X)$ and the digamma function $\psi(z) = d\log \Gamma(z)/dz$. We also use the function $X_n(u)$ appearing in the Borel transform of the coupling-constant expansion of $D_{\alpha}$,
\begin{align}
  B_{D_{\alpha},\mathrm{OPE}}(p^2, u) =p^2 \sum_{n=0}^{\infty}\left(\frac{M^2}{p^2}\right)^n \left(\frac{\mu^2}{p^2}\right)^u X_n(u),\\
  X_n(u) = 4\pi \frac{(4u^2-2u-2n)\Gamma(-n-2u)}{\Gamma(n+1)\Gamma(2-2u-2n)}.
\end{align}
\footnotetext{Strictly speaking, this procedure gives the part of the OPE of $\rho$ that does not vanish under the dual transform~\cite{Hayashi2023,takaura2024}. In the integrand for $\rho$, $D_{\alpha}(k^2)$ cannot be replaced by $D_{\alpha,\mathrm{OPE}}(k^2)$ in the region where $k^2$ is not sufficiently large. However, the contribution from this region vanishes after the dual transform and can therefore be neglected. This remains true for $\tilde{S}_{\mathrm{DV}}$ and $\tilde{S}_{\mathrm{OPE-DV}}$, for which the integration contour of the dual transform is restricted.}
Having adopted this Borel representation, it is still difficult to extract the individual Wilson coefficients $d_{n,k}^{(\rho)}$ from $B_{\rho,\mathrm{OPE}}(p^2, u)$.\footnotemark{} Instead, we perform the dual transform directly on $B_{\rho,\mathrm{OPE}}(p^2, u)$ with respect to $r=1/p^2$ and then perform the inverse Borel transform with respect to $u$.
\footnotetext{This is because the additional contribution noted in the previous footnote and the IR divergence must be taken into account.}

In this way, we obtain the dual transform of the OPE.\footnotemark{}
The perturbative expansion of $\rho$ contains renormalon divergences, but its renormalons are canceled by the dual transform~\cite{Hayashi2023}.
\footnotetext{The inverse Borel transform requires a regularization of the Borel integral because of its non-convergent behavior at large $u$. We follow ref.~\cite{takaura2024} for this regularization.}

Furthermore, some care is required when applying the discussion in sections~\ref{sec:relation} and \ref{sec:improve} to $\rho$. The derivatives with respect to $p^2$ in its definition produce a simple pole at the threshold $r=-1/M_{\mathrm{gap}}^2$. Therefore, when matching the integration contour of the dual-transformed OPE to that of the Borel-transformed low-energy expansion, the contribution from the small circular contour around $r=-1/M_{\mathrm{gap}}^2$ cannot be neglected; see figure~\ref{figofboreltransform}. Taking this contribution into account, eq.~\eqref{eq:OPE-DV} is modified in the present case to
\begin{align}
  \tilde{\rho}_{\mathrm{exact}} = \tilde{\rho}_{\mathrm{OPE}} - \tilde{\rho}_{\mathrm{DV}} + A e^{-z/M_{\mathrm{gap}}^2}.\label{eq:Amari}
\end{align}
Here, we can obtain $A=2/(3\sqrt{3}\pi)$.\footnotemark{} 
\footnotetext{This contribution was derived in ref.~\cite{takaura2024} using a numerically checked ansatz. In appendix~\ref{sec:rho-threshold}, we give an alternative analytic derivation and discuss the role of the threshold singularity.}

In the case of $\rho$, since $c_0=0$, both Methods I and II can extract $c_{m\ge1}$ without knowing $A$, as discussed in appendix~\ref{sec:rho-withoutA}. In this section, however, we include the additional term explicitly for simplicity of the argument. The extraction formulas then become
\begin{align}
  \frac{m!}{2\pi i}\oint_{z=z_0 e^{i\theta}} \frac{dz}{z^{m+1}} F_m(z) \tilde{\rho}_{\mathrm{OPE}}(z) +A\left(\frac{-1}{M_{\mathrm{gap}}^2}\right)^m \simeq \frac{c_m}{M^{2m}},\label{eq:rhoIcorrected}\\
  \tilde{\rho}_{\mathrm{OPE-DV}}(z) +Ae^{-z/M_{\mathrm{gap}}^2} = \sum_{m=0}^{\infty} J'_m \frac{z^m}{m!},\qquad J'_m \simeq \frac{c_m}{M^{2m}},\label{eq:rhoIIcorrected}
\end{align}
where $j=0$ and $F_m(z) =\Gamma(m+1,-z/M_{\mathrm{gap}}^2)/\Gamma(m+1)$.

Although these formulas explicitly include $A$, this is not essential for either method: the prescriptions without $A$ approach the same results as the OPE approximation improves.\footnotemark{} For a fair comparison, we also apply the corresponding correction to the method of ref.~\cite{takaura2024} in the numerical calculations below; this correction was not included in the original work.

\footnotetext{This agreement is examined analytically and numerically in appendix~\ref{sec:rho-withoutA}. Recall that, due to the asymptotic nature of the dual-transformed OPE (discussed in section~\ref{sec:StildeASYM}), increasing $n_{\mathrm{max}}$ requires taking $z_0$ sufficiently large for Method I.}

\begin{figure}
  \centering
  \begin{minipage}[t]{0.49\textwidth}
    \centering
    \includegraphics[width=\linewidth]{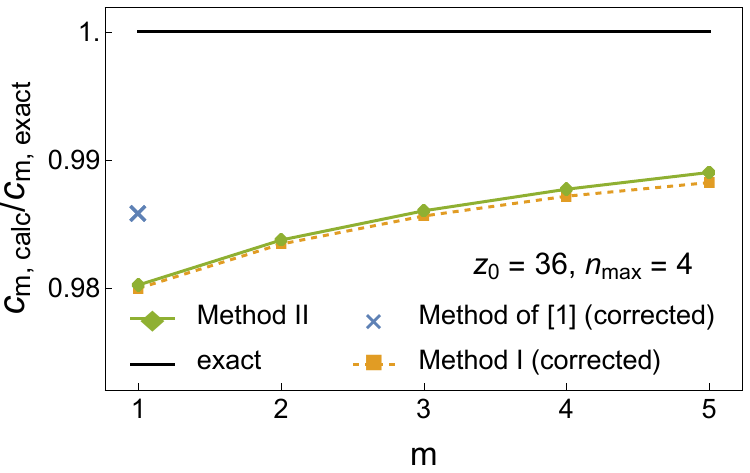}
    \par\smallskip
    \refstepcounter{jheppanel}\label{fig:rho-m}(\thejheppanel)
  \end{minipage}
  \hfill
  \begin{minipage}[t]{0.49\textwidth}
    \centering
    \includegraphics[width=\linewidth]{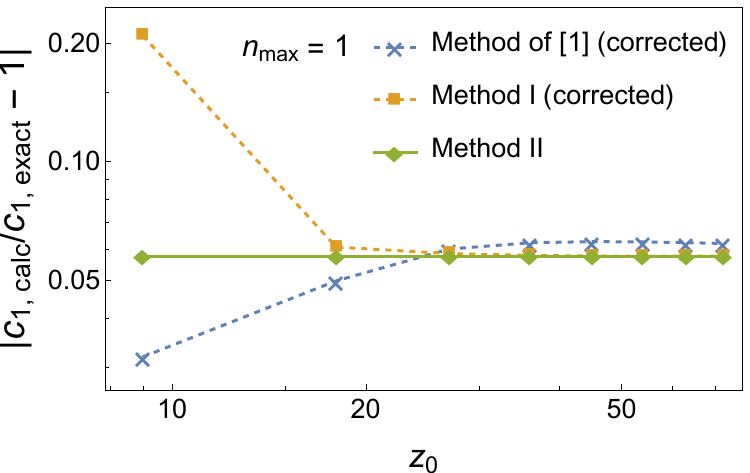}
    \par\smallskip
    \refstepcounter{jheppanel}\label{fig:rho-z0}(\thejheppanel)
  \end{minipage}
  \caption{Application of the methods to $\rho$: higher-order coefficients $c_m$ and $z_0$-dependence.
  (a) $c_m$ of $\rho$ calculated using Methods I and II and the method of ref.~\cite{takaura2024} (only for $c_1$), including a correction (see the main text), compared with the exact values. The vertical axis shows $c_{m,\mathrm{calc}}/c_{m,\mathrm{exact}}$. $n_{\mathrm{max}}=4$ and $z_0=36$ for Method I and the method of ref.~\cite{takaura2024}.
  (b) Dependence of the relative error in $c_1$ of $\rho$ on $z_0$, comparing Method I and the method of ref.~\cite{takaura2024} with the $z_0$-independent result of Method II. $n_{\mathrm{max}}=1$. Both axes are logarithmic.}
  \label{fig:rho-results1}
\end{figure}

\begin{figure}
  \centering
  \begin{minipage}[t]{0.49\textwidth}
    \centering
    \includegraphics[width=\linewidth]{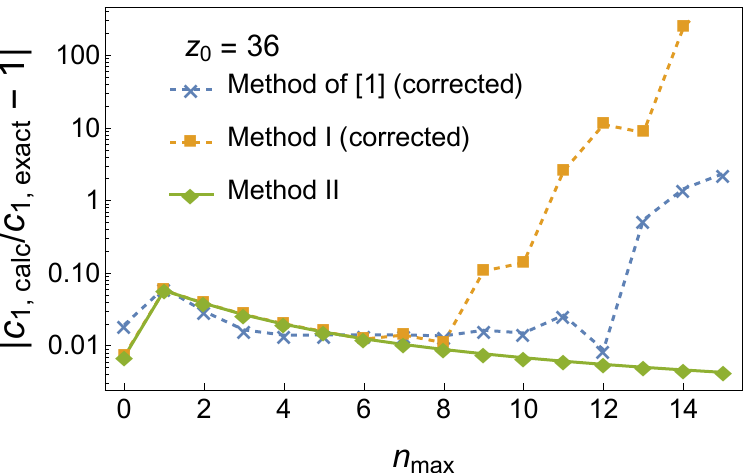}
    \par\smallskip
    \refstepcounter{jheppanel}\label{fig:rho-nmax}(\thejheppanel)
  \end{minipage}
  \hfill
  \begin{minipage}[t]{0.49\textwidth}
    \centering
    \includegraphics[width=\linewidth]{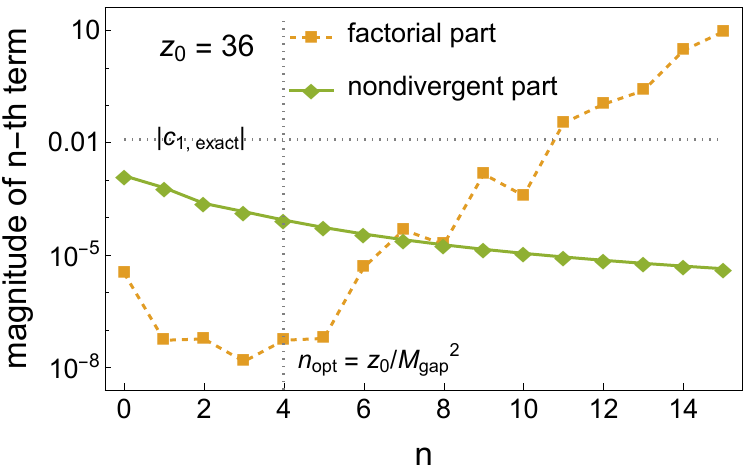}
    \par\smallskip
    \refstepcounter{jheppanel}\label{fig:rho-nmax-fact}(\thejheppanel)
  \end{minipage}
  \caption{Application of the methods to $\rho$: dependence on the OPE truncation order.
  (a) Dependence of the relative error in $c_1$ of $\rho$ on the OPE truncation order $n_{\mathrm{max}}$, comparing the convergence of the three methods. $z_0=36$ for Method I and the method of ref.~\cite{takaura2024}. The vertical axis is logarithmic.
  (b) Magnitudes of the $n$th-order contributions to $c_1$ on a logarithmic scale. The orange line shows the difference between the contributions from Methods I and II, while the green line shows the Method II contribution. The reference lines indicate $n=z_0/M_{\mathrm{gap}}^2=4$ and $|c_{1,\mathrm{exact}}|$.}
  \label{fig:rho-results2}
\end{figure}

The numerical results are shown in figures~\ref{fig:rho-results1} and~\ref{fig:rho-results2} and table~\ref{tab:rho-massgap}. Both Methods I and II successfully extract higher-order $c_m$, as in the $D_{\alpha}$ case.

In figure~\ref{fig:rho-results1}(\ref{fig:rho-m}), the accuracy of both Methods I and II appears to improve with increasing $m$, unlike that for $D_{\alpha}$. This is because the endpoint contribution $A(-1/M_{\mathrm{gap}}^2)^m$, which is included exactly in both methods, becomes more dominant at higher orders. In fact, the endpoint contribution alone already reproduces the exact coefficient $c_1$ with a relative error of about $10\%$. Consequently, errors in the contribution extracted from the OPE, arising from OPE truncation and, in Method I, finite $z_0$, become less significant relative to $c_m$.

The results of Methods I and II nearly overlap in figure~\ref{fig:rho-results1}(\ref{fig:rho-m}). This can be understood from figure~\ref{fig:rho-results1}(\ref{fig:rho-z0}), which shows that Method I rapidly approaches Method II as $z_0$ increases. Nevertheless, the small difference between the two methods increases with $m$, as in the $D_{\alpha}$ case.

Figures~\ref{fig:rho-results2}(\ref{fig:rho-nmax}) and \ref{fig:rho-results2}(\ref{fig:rho-nmax-fact}) also show behavior consistent with the estimate of eq.~\eqref{OPEculc}, as in the case of $D_{\alpha}$. The factorially growing part reaches its minimum near $n\sim z_0/M_{\mathrm{gap}}^2$, but the divergence in the extracted $c_1$ becomes visible only when this contribution becomes comparable in magnitude to $c_1$ itself.

Table~\ref{tab:rho-massgap} shows that the relative errors of the extracted $c_1$ remain below $10\%$ when the input value of $M_{\mathrm{gap}}^2$ is varied by $10\%$ from its exact value. For larger variations, such as doubling or halving $M_{\mathrm{gap}}^2$, the result is more sensitive than in the $D_{\alpha}$ case.

\begin{table}
  \centering
  \begin{tabular}{c|ccccc}
    \hline
    $M_{\mathrm{gap}}^2$ & $4.5$ & $8.1$ & $9\,\text{(exact)}$ & $9.9$ & $18$ \\
    \hline
    Method I  & $-19\%$ & $+6.4\%$ & $-2.0\%$ & $-9.6\%$ & $-49\%$ \\
    Method II & $-19\%$ & $+6.5\%$ & $-2.0\%$ & $-9.6\%$ & $-48\%$ \\
    \hline
  \end{tabular}
  \caption{Relative errors in the extracted $c_1$ of $\rho$ for different input values of $M_{\mathrm{gap}}^2$. $z_0=36$ and $n_{\mathrm{max}}=4$.}
  \label{tab:rho-massgap}
\end{table}

\FloatBarrier
\section{Conclusions and discussion}

In this work, we proposed two methods for extracting low-energy expansion coefficients at arbitrary orders from the OPE of one-scale observables in asymptotically free theories with a single threshold. The OPE is a high-energy representation constructed from Wilson coefficients obtained in ordinary perturbation theory and non-perturbative operator matrix elements that can be determined from experiments or lattice calculations. Our results thus demonstrate that, for this class of observables, low-energy information can be extracted from a representation originally formulated for high energies.

In ref.~\cite{takaura2024}, only the first nonzero coefficient of low-energy expansion was extracted. As an extension of this approach, we proposed two methods. In Method I, we generalized the weight function to eliminate the contribution from each term of the asymptotic expansion of the DV term. In Method II, we instead directly construct the dual-transformed OPE with the DV contribution subtracted. We showed that both methods allow the coefficients of the low-energy expansion to be determined to arbitrary order in $p^2/M^2$.

In particular, Method II avoids the optimal-truncation limitation caused by the additional $n!$-type growth introduced by the dual transform, which affects both Method I and the method of ref.~\cite{takaura2024}. Therefore, the accuracy can be improved simply by including higher-order terms in the OPE.

For $D_{\alpha}$, we derived a representation of the dual-transformed OPE directly in terms of the perturbative coefficients $d_{n,k}$, without using the Borel transform with respect to the coupling constant employed in ref.~\cite{takaura2024}. This representation makes the characteristic behaviors of the dual-transformed OPE and the DV term, $\tilde{S}_{\mathrm{OPE}} \sim \sum_n n! (M_{\mathrm{gap}}^2/z)^n$ and $\tilde{S}_{\mathrm{DV}}\sim e^{-z/M_{\mathrm{gap}}^2} \times (\text{inverse powers of}\ z)$ more transparent. This suggests that a similar construction may be possible for more general observables. The dominant structure is expected to remain unchanged when the $n$-dependence induced by anomalous dimensions is subleading compared with $n$, as discussed in appendix~\ref{sec:anomalous}.

Our results further support the idea proposed in ref.~\cite{takaura2024} that the OPE, although valid at high energies, contains information about the low-energy region. Extracting the low-energy expansion coefficients at arbitrary orders substantially increases the amount of low-energy information accessible from the OPE.

We point out a possible direction for extending the applicability of the OPE beyond its original domain. The dual transform has an inverse transform~\cite{Hayashi2023},
\begin{align}
S\left(p^2=\frac{1}{r}\right) &= r \int_{0}^{\infty} dz e^{-rz}\tilde{S}(z).
\end{align}
Applying this inverse transform to $\tilde{S}_{\mathrm{OPE-DV}}(z)$ defines a quantity $S_{\mathrm{OPE-DV}}(p^2)$. Since $\tilde{S}_{\mathrm{OPE-DV}}$ is constructed from the high-energy OPE by removing the contribution from the region where the OPE is unreliable, its inverse transform may provide a representation of $S(p^2)$ valid over the entire range of $p^2$. However, directly applying the inverse transform to the power expansion in $z$ in eq.~\eqref{OPE-DVculc} simply reproduces the low-energy expansion, which converges only for $|p^2|<M_{\mathrm{gap}}^2$. To obtain a representation valid over the entire range of $p^2$, the inverse transform must instead be applied before expanding in powers of $z$.

A remaining problem is the application to the case with multiple particle thresholds. As already pointed out in ref.~\cite{takaura2024}, realistic theories such as QCD contain multiple thresholds, extending to arbitrarily high energies. This problem is not resolved by the extensions developed in this work. The difficulty appears to originate from the fact that information associated with different thresholds cannot be separated by the contour structure of the dual transform alone. Both Method I (and the method of ref.~\cite{takaura2024}) and Method II suppress or exclude the contribution from the DV region of the dual transform, $-\infty<r<-1/M_{\mathrm{gap}}^2$. This treatment does not distinguish the effects of different thresholds and therefore cannot be directly applied even in the simple case of two thresholds, $M_{1,\mathrm{gap}}<M_{2,\mathrm{gap}}$. This remains an important subject for future work.

\appendix
\section{Derivation of \texorpdfstring{$\tilde{S}_{\mathrm{OPE}}, \tilde{S}_{\mathrm{DV}}, \tilde{S}_{\mathrm{OPE-DV}}$}{Stilde}}\label{sec:Appdual}

We use the notation and definitions introduced in the main text.

In many practical cases, $\langle \mathcal{O}_0 \rangle = 1$. In sections~\ref{sec:AppDerivOPE}--\ref{sec:AppDerivOPE-DV}, we neglect anomalous dimensions and assume that $\langle \mathcal{O}_n \rangle$ is independent of $\mu^2$, as in the case of $D_{\alpha}$, where $\langle \mathcal{O}_n \rangle = M^{2n}$. The effects of anomalous dimensions are discussed separately in section~\ref{sec:anomalous}.

Using the RG equation, each Wilson coefficient has the formal solution
\begin{align}
  C_n(p^2,\mu^2) = \left(\frac{\mu^2}{p^2}\right)^{\hat{H}} \sum_k d_{n,k}(0) g(\mu^2)^{k+1}, \label{eq:WilsonRGH}
\end{align}
where $\hat{H} = -\beta(g) \partial_g , \beta(g) = \mu^2 dg/d\mu^2$. 
On the other hand, since the Wilson coefficients are independent of the renormalization scale $\mu$, 
\begin{align}
  C_n(p^2,\mu^2) = C_n(p^2,\mu^2=p^2) = \sum_k d_{n,k}(0) g(p^2)^{k+1},
\end{align}
where
\begin{align}
  \left(\frac{\mu^2}{p^2}\right)^{\hat{H}} g(\mu^2)^{k+1} = g(p^2)^{k+1}. \label{eq:grunbyHwithoutanodim}
\end{align}
For simplicity, we write $d_{n,k}(0) = d_{n,k}$ hereafter.

\subsection{Derivation of \texorpdfstring{$\tilde{S}_{\mathrm{OPE}}$}{StildeOPE}}\label{sec:AppDerivOPE}

We substitute eq.~\eqref{eq:WilsonRGH} into the definition of the dual transform, eq.~\eqref{eq:dual}. Using the Hankel integral representation of the inverse Gamma function, we formally obtain
\begin{align}
  \tilde{S}_{\mathrm{OPE}}(z) = \sum_{n,k} d_{n,k}\langle \mathcal{O}_n \rangle z^{j-n} \left(\frac{\mu^2}{z}\right)^{\hat{H}} \frac{1}{\Gamma(1+j-n-\hat{H})} g(\mu^2)^{k+1}. \label{eq:StildeOPEfull}
\end{align}
Using the reflection formula and the asymptotic expansion of the Gamma function, we can formally write, for $n \gg 1$,
\begin{align}
  \frac{1}{\Gamma(1+j-n-\hat{H})} \sim (-1)^{n-j} \frac{\sin(\pi\hat{H})}{\pi} \Gamma(n-j) (n-j)^{\hat{H}} \left(1+\frac{\hat{H}(\hat{H}-1)}{2(n-j)} + \dots\right),
\end{align}
and we obtain 
\begin{align}
  \tilde{S}_{\mathrm{OPE}}(z) &\sim \sum_{n,k} d_{n,k}\langle \mathcal{O}_n\rangle z^{-n+j} \Gamma(n-j) L_{n,z}^{(k)},\\
  L_{n,z}^{(k)}&=(-1)^{n-j}\left(\frac{(n-j)\mu^2}{z}\right)^{\hat{H}} \frac{\sin(\pi\hat{H})}{\pi} (1+O(1/n))g(\mu^2)^{k+1}.
\end{align}
Using RG invariance, eq.~\eqref{eq:grunbyHwithoutanodim} allows us to rewrite $((n-j)\mu^2/z)^{\hat{H}}g(\mu^2)^{k+1}$ as $g(z/(n-j))^{k+1}$. Thus, the $n$-dependence of $L_{n,z}^{(k)}$ consists only of logarithmic factors, inverse powers, and a sign factor, introducing no additional factorial or exponential growth. Furthermore, from the expectation that the OPE converges in $|p^2| >M_{\mathrm{gap}}^2$, we can estimate $\sum_k d_{n,k} g^{k+1} \langle \mathcal{O}_n \rangle \sim M_{\mathrm{gap}}^{2n}$. Therefore,
\begin{align}
  \tilde{S}_{\mathrm{OPE}}(z) \sim \sum_{n} \left(\frac{M_{\mathrm{gap}}^2}{z}\right)^{n} \Gamma(n-j) L_{n,z}. \label{eq:AppOPEdual}
\end{align}

This establishes eq.~\eqref{OPEculc} under the assumptions of section~\ref{sec:review}, in particular for scale-independent non-perturbative matrix elements and a running coupling constant evaluated at any fixed finite order.

\subsection{Derivation of \texorpdfstring{$\tilde{S}_{\mathrm{DV}}$}{StildeDV}}\label{sec:AppDerivDV}
Substituting \eqref{eq:WilsonRGH} into the definition \eqref{tildeSDV}, we obtain
  \begin{align}
    \tilde{S}_{\mathrm{DV}}(z) =\sum_{n,k} \frac{d_{n,k}\langle \mathcal{O}_n \rangle}{\pi} z^{-n+j} \left(\frac{\mu^2}{z}\right)^{\hat{H}} \sin(\pi(\hat{H}+n-j)) \Gamma(\hat{H}+n-j, z/M_{\mathrm{gap}}^2) g(\mu^2)^{k+1}. \label{eq:StildeDVfull}
  \end{align}
  For fixed $n$ and $z/M_{\mathrm{gap}}^2 \gg 1$, we formally apply the asymptotic expansion of the incomplete Gamma function
  \begin{align}
    \Gamma\left(\hat{H}+n-j,\frac{z}{M_{\mathrm{gap}}^2}\right) \sim \left(\frac{z}{M_{\mathrm{gap}}^2}\right)^{\hat{H}+n-j-1} e^{\frac{-z}{M_{\mathrm{gap}}^2}} \left\{ 1+ \sum_{l=1}^{\infty} \left(\frac{M_{\mathrm{gap}}^2}{z}\right)^l \prod_{L=1}^{l} (\hat{H}+n-j-L)\right\}.
  \end{align}
Using the RG relation eq.~\eqref{eq:grunbyHwithoutanodim}, we obtain
\begin{align}
  \tilde{S}_{\mathrm{DV}}(z) \sim\, &e^{\frac{-z}{M_{\mathrm{gap}}^2}} \sum_{l=1}^{\infty} \left(\frac{M_{\mathrm{gap}}^2}{z}\right)^l  \notag\\
  &\times \left\{\sum_{n,k}\frac{(-1)^{n-j}d_{n,k}\langle \mathcal{O}_n \rangle}{\pi M_{\mathrm{gap}}^{2(n-j)}} \prod_{L=1}^{l-1} (\hat{H}+n-j-L)\sin(\pi \hat{H}) g(M_{\mathrm{gap}}^2)^{k+1} \right\}.
\end{align}
The expression inside the braces $\{\dots\}$ is independent of $z$. Therefore, the remaining $z$ dependence is entirely contained in the exponential factor $e^{-z/M_{\mathrm{gap}}^2}$ and the inverse powers $z^{-1}, z^{-2}, \ldots$, giving the asymptotic form~\eqref{SDVasym}.

\subsection{Derivation of \texorpdfstring{$\tilde{S}_{\mathrm{OPE-DV}}$}{StildeOPE-DV}}\label{sec:AppDerivOPE-DV}
Similarly, substituting \eqref{eq:WilsonRGH} into \eqref{StildeOPE-DV}, we obtain
  \begin{align}
    \tilde{S}_{\mathrm{OPE-DV}}(z) = \sum_{n,k} \frac{d_{n,k}\langle \mathcal{O}_n\rangle}{\pi} (-1)^{n-j}z^{j-n}\left(\frac{\mu^2}{z}\right)^{\hat{H}} \sin(\pi \hat{H})\gamma(\hat{H}+n-j,z/M_{\mathrm{gap}}^2) g(\mu^2)^{k+1}. \label{eq:StildeOPEDVfull}
  \end{align}
Using the series expansion of the lower incomplete Gamma function, 
  \begin{align}
    \gamma\left(\hat{H}+n-j,\frac{z}{M_{\mathrm{gap}}^2}\right) = \sum_{l=0}^{\infty} \frac{(-1)^{l}}{l! (\hat{H}+n-j+l)}\left(\frac{z}{M_{\mathrm{gap}}^2}\right)^{\hat{H}+n-j+l}.
  \end{align}
Using the RG relation eq.~\eqref{eq:grunbyHwithoutanodim}, we obtain
  \begin{align}
    \tilde{S}_{\mathrm{OPE-DV}}(z) = \sum_{l=0}^{\infty}\frac{z^l}{l!} \left\{\sum_{n,k} d_{n,k}\langle \mathcal{O}_n \rangle M_{\mathrm{gap}}^{2(-n+j-l)} \operatorname{sinc} \left(\pi(\hat{H}+n-j+l)\right) g(M_{\mathrm{gap}}^2)^{k+1} \right\}.
  \end{align}
  Here, $\operatorname{sinc}(x)=\sin(x)/x$ is an entire function. Since the expression inside $\{\dots\}$ is independent of $z$, we obtain \eqref{OPE-DVculc} and 
  \begin{align}
    J_m = \left\{\sum_{n,k} d_{n,k}\langle \mathcal{O}_n \rangle M_{\mathrm{gap}}^{2(-n+j-m)} \operatorname{sinc} \left(\pi(\hat{H}+n-j+m)\right) g(M_{\mathrm{gap}}^2)^{k+1} \right\}.\label{eq:explicitJm}
  \end{align}
  Thus, $\tilde{S}_{\mathrm{OPE-DV}}(z)$ consists only of nonnegative powers of $z$. Moreover, unlike $\tilde{S}_{\mathrm{OPE}}$, the transformation introduces no additional factorial factor $n!$.

\subsection{Effects of anomalous dimensions} \label{sec:anomalous}

In general, operator matrix elements can have anomalous dimensions. The formal solutions for the Wilson coefficients then become
\begin{align}
C_n(p^2,\mu^2) = \left(\frac{\mu^2}{p^2}\right)^{\hat{H}'_n} \sum_k d_{n,k}g(\mu^2)^{k+1}.  
\end{align}
Here, $\hat{H}'_n=\hat{H}+\gamma_n(g),\quad \mu^2 d/d\mu^2 \langle \mathcal{O}_n\rangle_{\mu^2} =-\gamma_n(g)\langle \mathcal{O}_n\rangle_{\mu^2}$ and
\begin{align}
  \left(\frac{\mu^2}{p^2}\right)^{\hat{H}'_n} g(\mu^2)^{k+1} \langle \mathcal{O}_n\rangle_{\mu^2} = g(p^2)^{k+1} \langle \mathcal{O}_n\rangle_{p^2}.  \label{eq:grunbyHdashwithanodim}
\end{align}
The above arguments can be extended to the case with anomalous dimensions, and the conclusions remain essentially unchanged.
For $\tilde{S}_{\mathrm{OPE}}$, the factorial large-$n$ behavior originates from $1/\Gamma(1+j-n-\hat{H})$. Therefore, provided that the $n$-dependence induced by $\hat{H} \to \hat{H}'_n$ is subleading compared with $n$, the leading $n!$ behavior is unchanged.

For $\tilde{S}_{\mathrm{DV}}$ and $\tilde{S}_{\mathrm{OPE-DV}}$, the RG relation \eqref{eq:grunbyHwithoutanodim} allows the $z$-dependence associated with the running coupling to be absorbed into quantities evaluated at fixed scale $M_{\mathrm{gap}}^2$. The same mechanism works in the presence of anomalous dimensions through $\hat{H}'_n$ and \eqref{eq:grunbyHdashwithanodim}. Consequently, the remaining $z$-dependence retains the same form as in the case without anomalous dimensions: $e^{-z/M_{\mathrm{gap}}^2}$ times inverse powers of $z$ for $\tilde{S}_{\mathrm{DV}}(z)$, and only nonnegative powers of $z$ for $\tilde{S}_{\mathrm{OPE-DV}}(z)$.

\section{Additional contribution at threshold and its treatment for \texorpdfstring{$\rho$}{rho}}

\subsection{Contribution from a contour around threshold singularity}\label{sec:rho-threshold}

The third term on the right-hand side of eq.~\eqref{eq:Amari} arises because of the definition $\rho(p^2) =p^2 (\frac{d}{dp^2})^2 \Pi_{\sigma}(p^2)$. This term corresponds to the contribution from the contour integral near $r=-1/M_{\mathrm{gap}}^2$ that remains when comparing the dual transform of the OPE and low-energy expansion. Ref.~\cite{takaura2024} showed that its form can be anticipated and determined numerically. Here, we show that it can also be derived analytically.

As illustrated in figure~\ref{figofboreltransform}, integrating along the small circular contour $r=-1/M_{\mathrm{gap}}^2 + \epsilon e^{i \theta}, -\pi<\theta<\pi$ gives
\begin{align}
  \tilde{\rho}_{\mathrm{th}} =A e^{-z/M_{\mathrm{gap}}^2}.
\end{align}
To derive this, it is sufficient to show that the leading singular behavior is 
\begin{align}
  \left(\frac{d}{dp^2}\right)^2 \Pi_{\sigma}(p^2) \simeq \frac{-A}{p^2+M_{\mathrm{gap}}^2},
\end{align}
where $\simeq$ denotes equality of the leading singular term. Regular and less singular terms at $p^2 = -M_{\mathrm{gap}}^2$ can be neglected, as they vanish in the contour integral.

To evaluate $\Pi_{\sigma}(p^2)$ near the threshold $p^2 = -M_{\mathrm{gap}}^2$, we apply the threshold expansion~\cite{Beneke_1998}. The dominant contribution comes from the region where the internal $\sigma$ propagator, $1/((p+k)^2+M^2)$, is near its mass shell and the $\delta\alpha$ propagator, $D_\alpha(k^2)$, is near its threshold. In this region, the corresponding energies are close to $M$ and $2M$, respectively, allowing a nonrelativistic expansion. We can approximate
\begin{align}
  D_{\alpha}(k^2\sim -4M^2) &\sim 8M\sqrt{k^2+4M^2},\\
  \Pi''_{\sigma}(p^2\sim -9M^2) &\simeq \frac{4M}{\pi^2} \int dk^0\, dk^1\, \frac{(p\cdot(p+k))^2 \sqrt{k^2+4M^2}}{(p^2)^2 ((p+k)^2+M^2)^3}. 
\end{align}
Using Minkowski momentum $p_M$, with $p^2=-p_M^2$, and performing the $k^0$ integration while carefully accounting for the locations of the singularities, we obtain
\begin{align}
  \Pi''_{\sigma}(p_M^2\sim 9M^2) &\simeq \frac{-M^{-3/2}}{36\pi} \int dk^1\, \left\{ \left(3M-p_M^0+\frac{(p_M^1)^2}{6M}\right) + \frac{3}{4M}(k^1+\frac{2p_M^1}{3})^2\right\}^{-3/2}\\
  &\sim \frac{2}{3\sqrt{3}\pi}\frac{1}{p_M^2-9M^2}.
\end{align}
Therefore, $A=\frac{2}{3\sqrt{3}\pi} \simeq 0.1225$.

In the case of $\rho$, the derivatives introduced to render the observable UV finite promote the threshold singularity into a simple pole, making the endpoint contribution non-negligible. Although we used the exact expression for $D_\alpha$ to determine $A$ in the present model, the full exact expression is not necessarily required.

The required treatment depends on the strength of the threshold singularity. If it is weaker than a simple pole, the contribution from the small circular contour around the branch point vanishes as its radius approaches zero, and the discussion in section~\ref{sec:improve} applies without modification. A simple-pole singularity requires a correction analogous to that for $\rho$, whereas a stronger singularity is expected to require further modifications.

\subsection{Extraction without the threshold contribution}\label{sec:rho-withoutA}

In the case of $\rho$, the low-energy expansion coefficients $c_{m\ge1}$ can be extracted without knowing the value of $A$, by using the weight function $F_{m-1}(z)$ instead of $F_m(z)$ in Method I or using $-J_0$ as an approximation of $A$ in Method II. Furthermore, as the OPE approximation becomes more accurate, the $c_m$ obtained without using $A$ approaches the result obtained with the explicit $A$ correction.

In Method I, the threshold contribution can be eliminated together with the DV contribution by an appropriate choice of the weight function. The additional term $Ae^{-z/M_{\mathrm{gap}}^2}$ can be regarded as the $l=0$ term of the asymptotic expansion of $\tilde{\rho}_{\mathrm{DV}}$ in eq.~\eqref{SDVasym}. Thus, the DV-vanishing condition~\eqref{eq:FcondDVint} must also be imposed for $l=0$, in addition to $l\ge1$.

On the other hand, $c_0 =0$ due to the derivatives in the definition of $\rho$ in eq.~\eqref{eq:rho}. Therefore, the $c_m$-extracting condition \eqref{eq:condLEE} can be relaxed to allow a contribution proportional to $c_0$. Using this freedom, we can also eliminate the $l=0$ term by choosing the weight function $F_{m-1}$:
\begin{align}
  F_{m-1}(z) = \frac{\Gamma(m,\frac{-z}{M_{\mathrm{gap}}^2})}{\Gamma(m)}.\label{eq:F_mrho}
\end{align}
With this choice, the extraction formula is modified to 
\begin{align}
    \frac{m!}{2\pi i}\oint_{z=z_0 e^{i\theta}} \frac{dz}{z^{m+1}} F_{m-1}(z)\tilde{\rho}_{\mathrm{OPE}}(z) \simeq \frac{c_m}{M^{2m}} - \frac{c_0}{M^{0}} \left(\frac{-1}{M_{\mathrm{gap}}^2}\right)^m. \label{eq:rhoextractwithoutA}
\end{align}
Therefore, since $c_0=0$ in the case of $\rho$, $c_m$ can be extracted without using $A$.

In Method II, the contribution of the threshold term to $c_m$ is given by $A(-1/M_{\mathrm{gap}}^2)^m$, as follows from eqs.~\eqref{OPE-DVculc}, \eqref{eq:Amari}, and \eqref{eq:rhoIIcorrected}. Since the exact coefficient $c_0$ vanishes, we have
\begin{align}
  J'_0 = J_0 + A \simeq 0,
  \qquad \therefore \quad A \simeq -J_0.
\end{align}
Substituting this relation into the extraction formula, we obtain
\begin{align}
  \frac{c_m}{M^{2m}} \simeq J_m - J_0\left(\frac{-1}{M_{\mathrm{gap}}^2}\right)^m. \label{eq:rhoIIwithoutA}
\end{align}
Thus, $c_m$ can also be extracted without using $A$ in Method II.

In practice, however, the left-hand side of eq.~\eqref{eq:rhoextractwithoutA} and $J_0$ in eq.~\eqref{eq:rhoIIwithoutA} are evaluated using the OPE truncated at finite order. As seen from eq.~\eqref{eq:Amari}, the exact relation $c_0=0$ results from the cancellation between the $z^0$ coefficients of $\tilde{\rho}_{\mathrm{OPE}}-\tilde{\rho}_{\mathrm{DV}}$ and $Ae^{-z/M_{\mathrm{gap}}^2}$. This cancellation is incomplete at finite OPE truncation, leaving a residual contribution in both methods.

\begin{figure}[!tbp]
  \centering
  \includegraphics[width=0.6\textwidth]{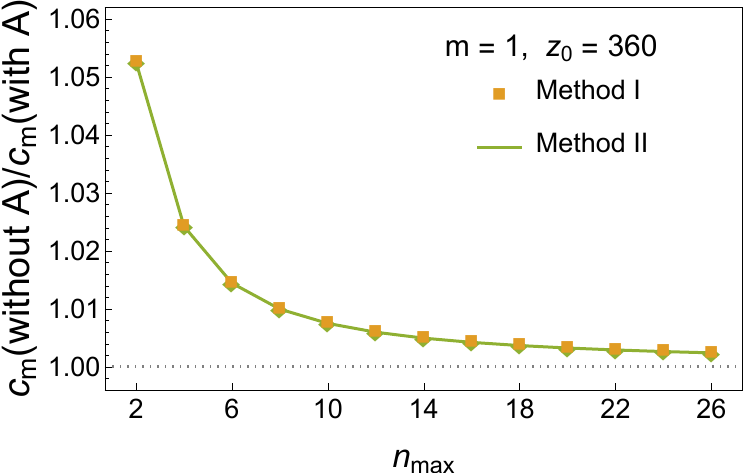}
\caption{
Ratio of $c_1$ calculated without using $A$ to that calculated with the explicit $A$ correction, as a function of $n_{\mathrm{max}}$, for Methods I and II. In Method I, the numerator and denominator are calculated using $F_{m-1}$ and $F_m$, respectively (eqs.~\eqref{eq:rhoextractwithoutA} and \eqref{eq:rhoIcorrected}). In Method II, $-J_0$ is used instead of $A$ for the numerator (eqs.~\eqref{eq:rhoIIwithoutA} and \eqref{eq:rhoIIcorrected}). For Method I, $z_0=360$ is fixed. Both ratios approach $1$ as $n_{\mathrm{max}}$ increases. Since the two results nearly coincide, Method I is shown with markers and Method II with a solid line for clarity.
}
  \label{fig:rho_ratiowithandwithoutA}
\end{figure}

In Method I, the exact value $c_0=0$ in eq.~\eqref{eq:rhoextractwithoutA} is effectively replaced by a nonzero residual $\delta c_0^{(n_{\mathrm{max}})}$ due to finite OPE truncation. Since this residual $\delta c_0^{(n_{\mathrm{max}})}$ originates from the incomplete cancellation at finite OPE truncation, it decreases as the OPE approximation becomes more accurate. In fact, the extraction formulas with and without the explicit $A$ term are related by
\begin{align}
  \frac{c_m^{(n_{\mathrm{max}})}}{M^{2m}} &= \frac{m!}{2\pi i}\oint_{z=z_0 e^{i\theta}} \frac{dz}{z^{m+1}} F_m(z) \tilde{\rho}_{\mathrm{OPE}}^{(n_{\mathrm{max}})}(z) +A\left(\frac{-1}{M_{\mathrm{gap}}^2}\right)^m \\
  &= \frac{m!}{2\pi i}\oint_{z=z_0 e^{i\theta}} \frac{dz}{z^{m+1}} F_{m-1}(z) \tilde{\rho}_{\mathrm{OPE}}^{(n_{\mathrm{max}})}(z) +\left(\frac{-1}{M_{\mathrm{gap}}^2}\right)^m \delta c_0^{(n_{\mathrm{max}})},\label{eq:rhoc0=!0correction}
\end{align}
where
\begin{align}
  \delta c_0^{(n_{\mathrm{max}})} = \frac{1}{2\pi i} \oint \frac{dz}{z} F_0(z) \tilde{\rho}_{\mathrm{OPE}}^{(n_{\mathrm{max}})}(z) +A.\label{eq:rhoc0=!0}
\end{align}
Thus, the two extraction formulas for Method I, \eqref{eq:rhoIcorrected} and \eqref{eq:rhoextractwithoutA} (with $c_0=0$), agree as $\delta c_0^{(n_{\mathrm{max}})}$ approaches zero. Recall that, due to the asymptotic nature of the dual-transformed OPE, increasing $n_{\mathrm{max}}$ requires sufficiently large $z_0$. The Method I result in figure~\ref{fig:rho_ratiowithandwithoutA} demonstrates this agreement numerically. 

For the method of ref.~\cite{takaura2024}, the finite-truncation correction can be implemented analogously to eqs.~\eqref{eq:rhoc0=!0correction} and \eqref{eq:rhoc0=!0}, using its weight function $F_{(0)}$ to evaluate the residual $\delta c_0^{(n_{\mathrm{max}})}$ contribution. We have also confirmed that the $c_1$ obtained without correcting the residual $c_0$ contribution approaches the corrected result as $n_{\mathrm{max}}$ increases.

Similarly, in Method II, $J_0^{(n_{\mathrm{max}})}$ approximates $-A$ only up to finite-truncation effects. However, since Method II does not involve the factorial enhancement associated with the dual transform, this approximation is expected to improve as $n_{\mathrm{max}}$ increases. The Method II results in figure~\ref{fig:rho_ratiowithandwithoutA} demonstrate that the extraction without $A$ approaches that with the explicit $A$ correction. 

\FloatBarrier
\acknowledgments

The authors are particularly grateful to H. Takaura for valuable advice and fruitful discussions. We also thank M. Yamaguchi and A. Yoshino for discussions. H.F. acknowledges financial support from the Tohoku University Advanced Graduate School LEAP Program. The work of Y.S. was supported in part by JSPS KAKENHI Grant Numbers JP23K03404 and JP26K07098. H.F. used ChatGPT (OpenAI) for language editing and assistance with some calculations. All AI-assisted outputs used in this work were reviewed and verified by the authors.

\bibliographystyle{JHEP}
\bibliography{myreferences}

\end{document}